\documentclass[fleqn,usenatbib]{mnras}

\usepackage{newtxtext,newtxmath}

\usepackage[T1]{fontenc}

\DeclareRobustCommand{\VAN}[3]{#2}
\let\VANthebibliography\thebibliography
\def\thebibliography{\DeclareRobustCommand{\VAN}[3]{##3}\VANthebibliography}

\usepackage{graphicx}	
\usepackage{amsmath}	
\DeclareUnicodeCharacter{2212}{-}

\title[Evolution in filaments and nodes]{Evolution of HOD and galaxy properties in filaments and nodes of the cosmic web}

\author[Perez et al.]{
Noelia R. Perez,$^{1}$\thanks{E-mail: noeliarocioperez@gmail.com (NP)}
Luis A. Pereyra,$^{2,3}$
Georgina Coldwell,$^{1}$
Ignacio G. Alfaro$^{2,3}$
Facundo Rodriguez,$^{2,3}$
\newauthor and  Andr\'{e}s N. Ruiz$^{2,3}$
\\
$^{1}$Facultad de Ciencias Exactas, F\'{i}sicas y Naturales, Departamento de Geof\'{i}sica y Astronom\'{i}a, CONICET Universidad Nacional de San Juan, Av. Ignacio de la\\  Roza 590 (O), J5402DCS, Rivadavia, San Juan, Argentina\\
$^{2}$Instituto de Astronom\'{i}a Te\'{o}rica y Experimental, CONICET-UNC, Laprida 854, X5000BGR C\'{o}rdoba, Argentina\\
$^{3}$Observatorio Astron\'{o}mico de C\'{o}rdoba, UNC, Laprida 854, X5000BGR C\'{o}rdoba, Argentina
}

\date{Accepted XXX. Received YYY; in original form ZZZ}

\pubyear{2023}

\begin{document}
\label{firstpage}
\pagerange{\pageref{firstpage}--\pageref{lastpage}}
\maketitle

\begin{abstract}
We study the evolution of the Halo Occupation Distribution (HOD) and galaxy properties of nodes and filamentary structures obtained by \textsc{DisPerSE} from the \textsc{Illustris TNG300-1} hydrodynamical simulation, in the redshift range $0 \leq z \leq 2$. 
We compute the HOD in filaments and nodes and fit the HOD parameters to study their evolution for both faint and bright galaxies. In nodes, the number of faint galaxies increases with decreasing redshift in the low mass halos, while no significant differences are seen in high mass halos. 
Limiting the HOD to bright galaxies shows that halos increase in mass more than the number of bright galaxies they accrete. 
For filaments, no large differences in HOD are found for faint galaxies, although for brighter galaxies the behaviour is the same as in nodes. 
The HOD parametrization suggests that filaments have no effect on the mass required to host a galaxy (central or satellite), whereas nodes do.
The results of the study indicate that with this parametrization, filaments do not seem to affect the stellar mass content of galaxies. In contrast, nodes appear to affect halos with masses below approximately $10^{12.5} h^{-1} M_{\odot}$ at local redshift.
The analysis of the galaxy colour evolution shows a reddening towards lower redshift, although these processes seem to be more efficient in massive halos, with a strong effect on bright galaxies. 
The general evolution suggests that the building of galaxy population within halos is influenced by both the accretion of faint galaxies and the mass growth of the bright ones. 
\end{abstract}




\section{Introduction}

At the megaparsec scale, matter is not evenly distributed. Instead, it is found in a pattern called the Cosmic Web \citep{deLapparent1986,bond96}.
In this network, we can distinguish the largest
non-linear structures in the Universe, such as nodes, filaments, walls, and voids.
In the current cosmological model, this web pattern is a consequence of the growth of initial density perturbations \citep{bond96,Zel'dovich1970,Doroshkevich1980} due to gravitational instability in the early Universe \citep{Lifshitz46,Peebles1980}. 
Thus, cosmic structures bear an embryonic imprint in the primordial density field, indicating that the evolution of the initial density field peaks led to the large-scale structure currently observed \citep{Bond1996_Myers,Aycoberry2023}.

In this context, dark matter halos form from initial density perturbations in the density field of the early Universe and grow with time due to gravitational instability \citep{Lifshitz46,Peebles1980,shandarin89}. Baryonic matter follows the gravitational potential wells of dark matter, where it cools and forms visible systems, stars, and galaxies \citep{WhiteRees78,White1994}. Later structures then grow hierarchically, from smaller to larger ones, and halos containing large galaxies are formed through repeated mergers of smaller halos.

The formation and evolution of galaxies occur within dark matter halos, where the properties of the galaxies are influenced by a number of processes occurring within the halo itself. These include processes such as gas cooling, feedback, and mergers, as well as the properties of the halo itself \citep{Mo1998, VandenBosch1998, Cole2000}.
Then, the relationship between the galaxies and the dark matter halos cannot be precisely determined due to the large number of processes involved.
The Halo Occupation Distribution (HOD) is a powerful tool for inferring, on average for a galaxy population, the halo-galaxy relationship.
It outlines the relationship between galaxies and their host halo by the probability distribution that a halo of mass $M$ contains $N$ galaxies with specific characteristics. \citep{Peacock2000,Berlind2002,Berlind2003,Zheng2005,Guo2015,Rodriguez2020}.

The HOD characterisation suggests that the cosmological model determines the properties of the halo distribution, while the theory of galaxy formation specifies how these halos are populated \citep{Berlind2002}.
Additionally, the HOD framework assumes that the mass of the host halo is the primary factor influencing galaxy properties.
However, other factors such as formation time, spin, concentration, neighbour mass, and large-scale structure also impact the spatial distribution of galaxies 
\citep{Gao2005,Gao2007,Mao2018,Musso2018,Mansfield2020}.
Several works have used the HOD approach to constrain models of galaxy formation and evolution \citep{Berlind2003,Kravtsov2004,Zehavi2018}, and cosmological models \citep{vandenBosch2003,Zheng2007}.
In addition, the HOD has allowed the creation of mock catalogues.
As a result, \cite{Grieb2016} used the clustering statistics obtained from the HOD and its parametrization to populate the halos with galaxies, generating a mock catalogue.

The HOD framework has also been used with observational data.
\cite{Alam2020,Yuan2022b} developed a generalised HOD model to describe the populations of the extended Baryon Oscillation Spectroscopic Survey (eBOSS, \cite{Dawson2016}) tracers:
Luminous Red Galaxies (LRG), Emission Line Galaxies(ELG) and Quasi-Stellar Objects (QSO).
They found that for $0.7 \leq z \leq 1.1$, ELGs inhabit low mass halos, QSOs intermediate-mass halos, and LRGs high mass halos.
In addition, the HOD framework has been used to compare catalogues obtained from the Sloan Digital Sky Survey (SDSS) with mock catalogues \citep{Skibba&Sheth2009,Rodriguez2020}.
In this line, \cite{Rodriguez2020} developed an algorithm that combines two galaxy group identifiers: friend-of-friend \citep{Huchra1982} and halo-based \citep{Yang2005}. They applied this algorithm to the SDSS-DR12 spectroscopic galaxy catalogue \citep{Alam2015} and obtained two catalogues. Thus, the HOD was computed for these catalogues and a mock catalogue to determine the appropriateness of the group definition and the reliability of the mass estimate. Their study concluded that the identification method used did not have a significant impact on the measurement of HOD.

On the other hand, the large-scale environment affects the galaxy properties such as shape and spin alignment \citep{Aragon-Calvo_2007,Hahn2010,Tempel2013,Zhang2015,Ganeshaiah_Veena2018,Wang2020,Lee2023}, galaxy stellar mass, star formation rate and colour \citep{Einasto2008,Lietzen2012,Darvish2016,Malavasi2017,Kraljic2018,Laigle2018}.
Additionally, there is evidence that the HOD is impacted by large-scale neighbourhood \citep{Artale2018,Zehavi2018,Alfaro2020, Alfaro2021}.
Thus, the formation and evolution of galaxies is influenced by the tides of the large-scale environment. 
These physical properties depend on the scale and distance to the components of the cosmic network in a manner specific to each observable \citep{Kraljic2018}.

The evolution of the Cosmic Web has been extensively studied both theoretically and observationally.
Matter is transported from the less dense regions, such as voids, to the denser ones (i.e., galaxy clusters). Therefore, filaments serve as the channels through which matter flows from voids and walls to nodes. This flow can be observed by comparing the structures at $z=0$ with earlier epochs \citep{Aragon-Calvo2010}.
With redshift, the matter in the voids slowly decreases and their volume increases, the mass and volume of the walls decrease, the matter in the filaments concentrates into fewer structures that become more massive, and the nodes become more massive \citep{Cautun2014}. 

Furthermore, the characterisation of the cosmic web has become more accurate in recent years, allowing for more detailed analyses of the links between the properties of gas and galaxies and their specific locations within these structures.
In particular, the Discrete Persistent Structure Extractor \citep[DisPerSE;][]{Sousbie2011_a, Sousbie2011_b} provides a formalism that facilitates the detection of large-scale structures. 
\cite{MonteroDorta2023} used the structures identified by DisPerSE to examine the relationship between the distances separating these structures and the secondary dependencies at a fixed halo mass, also known as the halo assembly bias.

Also, in the context of HOD evolution, \cite{Zheng2005} proposed a parametrization of the halo occupation distribution based on five parameters. Later, based on this work, 
\cite{Contreras2023} studied the evolution of the HOD by evaluating how the parameters involved vary from $z=0$ to $z=3$. 
They found that cosmology governs the evolution of the halo occupation for stellar mass-selected samples.
Moreover, \cite{Contreras2017} found that these parameters are similar for two independent semi-analytical galaxy formation models. 

In \cite{Perez2024}, we explored the HODs and galaxy properties in nodes and filaments of the local Universe. We extend this investigation by studying the effect of the evolution of the cosmic web on the relationship between halos and galaxies through different redshifts. We then analyse the evolutionary trends in the distribution of halo and galaxy properties within both cosmic structures, from the local Universe to redshifts up to $z=2$.
This paper is structured as follows:
in Section \ref{sec:data} we summarise the data and the sample selection. In Section \ref{sec:analysis} we analyse the evolution of HOD in nodes and filamentary structures, the HOD parametrization and the evolution of galaxy properties. Finally, in Section \ref{sec:conclusions} we summarise and discuss the main results.

\section{Data} \label{sec:data}
\subsection{IllustrisTNG}

\textsc{IllustrisTNG} \footnote{https://www.tng-project.org/} 
\citep{Nelson2018,Pillepich2018_a,Pillepich2018_b} is a suite of cosmological hydrodynamic simulations developed from the original \textsc{Illustris} simulations \citep{Genel2014,Vogelsberger2014_a,Vogelsberger2014_b} and performed with the \textsc{arepo} moving mesh code \citep{Springel2010} in three simulation volume boxes: TNG50, TNG100 and TNG300.
The cosmological parameters are based on the Planck 2015 results  
\citep{planckresults}: $\Omega_{\text{m},0}=0.3089$, $\Omega_{\Lambda,0}= 0.6911$, $\Omega_{\text{b},0}= 0.0486$, $\text{h}= 0.6774$, $\text{n}_{\text{s}}= 0.9667$ and $\sigma_{8}= 0.8159$.
These simulations improve physical models of galaxies by accounting for the growth and feedback of supermassive black holes, galactic winds and stellar evolution, and the chemical enrichment of gas \citep{Pillepich2018_a}.

TNG300 has $2500^3$ dark matter particles and $2500^3$ gas particles with masses of $5.9 \times 10^7 \text{M}_{\odot}$ and $1.1 \times 10^7 \text{M}_{\odot}$ respectively, within a box of $205\ h^{-1} \text{Mpc}$.
In particular, we are working with the main high resolution run of TNG300, TNG300-1.
In TNG the halos (groups) are detected using the friend-of-friend algorithm \citep{Huchra1982} with linking length $b = 0.2$ and to identify subhalos (galaxies) the \textsc{subfind} algorithm \citep{Springel2001} is used.
There are 100 snapshots available from $z=0$ to $z=20.05$.

\subsection{DisPerSE}

\begin{table}
	\centering
	\caption{Percentage of halos and subhalos in each sample for the given redshift snapshots.}
	\label{tab:desc_samples}
	\begin{tabular}{ccccccc}
		\hline
        $z$ & \multicolumn{3}{|c|}{Halos}& \multicolumn{3}{|c|}{Subhalos} \\ \hline
         & All & Filaments & Nodes & All & Filaments & Nodes\\
         &     & (\%) & (\%) &  & (\%) & (\%)\\

		\hline
		0.0 & 212749 & 27.29 & 3.31 & 264407 & 27.70 & 30.71\\ 
		0.5 & 229835 & 25.89 & 3.08 & 232912 & 30.12 & 26.44\\
		1.0 & 232189 & 24.07 & 2.77 & 197254 & 30.71 & 22.41\\
        1.5 & 218291 & 22.63 & 2.42 & 146448 & 30.82 & 18.95\\
        2.0 & 190256 & 21.25 & 2.11 & 102788 & 30.88 & 15.42\\        
		\hline
	\end{tabular}
\end{table}

\textsc{DisPerSE} \footnote{http://www2.iap.fr/users/sousbie/web/html/index888d.html?archive} 
\citep[Discrete Persistent Structures Extractor,][]{Sousbie2011_a} is a multiscale identifier of topological features, based on the discrete Morse theory.  It is mainly used
to identify filamentary patterns, but can also identify nodes, walls and voids.

From a discrete set of points (in this case galaxies) and through the Delaunay Tessellation Field Estimator (DTFE, \citet{Schaap2000}), the algorithm generates a density field that allows the identification of critical points.
Taking into account mathematical properties that allow linking the topology and geometry of the density field, the critical points together with their ascending and descending manifolds $0$, $1$, $2$ and $3$ can be related to clusters, filaments, walls and voids respectively. 
Maxima and saddle points are associated with filaments of the cosmic web, which follow lines of constant gradient in the density field.  Each filament is composed of small segments, a filament being defined as the set of straight segments connecting a point of maximum density to a saddle of the density field. The importance of the topological features can be quantified using the persistence level to filter out the Poisson sampling noise and the intrinsic uncertainty within the data set.

For this work, we use the Cosmic Web Distances catalogue \citep{Duckworth2020a,Duckworth2020b} 
\footnote{https://github.com/illustristng/disperse\_TNG}
based on the \textsc{DisPerSE} code, and available for TNG300-1, which provides the cosmic web distances between each subhalo and the nearest cosmic web structures.
In building this catalogue, the authors have selected galaxies (subhalos) with masses greater than $M_* > 10^{8.5} h^{-1} M_{\odot}$ as points to define the cosmic web and they have used a conservative persistence level of $4\sigma$ to avoid spurious structure detections, in line with the works of \cite{Malavasi2020b} and \cite{GalarragaEspinosa2024}.
It is expected that a small cutoff could increase the number of filaments by including weak branches, albeit at the cost of introducing more noise, while at higher persistence values only large scale filaments would survive. 
Furthermore, the persistence level selected ensures that the number of critical points in the DisPerSE skeleton remains consistent on average across the redshift snapshots used in this study.
For a detailed analysis of the calibration of the DisPerSE parameters, see \cite{GalarragaEspinosa2024}.

\subsection{Sample selection}

In this paper we have used the TNG300-1 Groupcat in $5$ different snapshots:  $z=0.0, 0.5, 1, 1.5 \ \text{and} \ 2$, in order to analyse the evolution of the HOD. 
For consistency with the Cosmic Web Distances catalogue, we have selected subhalos with $\text{M}_* > 10^{8.5}\ h^{-1} \text{M}_{\odot}$ for each redshift, where $\text{M}_*$ is defined as the sum of the mass of the member stellar particles, restricted to stars within the radius at which the surface brightness profile drops below $20.7$ mag $\text{arcsec}^{−2}$ in the K band.
We also set these subhalos so that they belong to halos with masses $\text{M}_{200} \geq 10^{11}\ h^{-1} \text{M}_{\odot}$ to obtain well resolved halos with about $10^3$ particles. 
Here, ${\rm M}_{200}$ is the mass enclosed in a region of $200$ times the critical density.

We are particularly interested in determining the HOD in different cosmic web structures: filaments and nodes.
To achieve this, we define nodes as those halos whose distance to the peaks of the DisPerSE density field is less than $\rm R_{\rm 200}$ ($\text{d}_{\text{n}} \leq 1 \text{R}_{\text{200}}$), where $\rm R_{\rm 200}$ is the comoving radius of a sphere centred on the halo whose mean density is $200$ times the critical density of the Universe.
The $\rm R_{\rm 200}$ radius is commonly used to define the 'virial' radius by several authors, including \cite{Carlberg1997,Tinker2005,Rines2010,Pizzardo2023}, etc. This radius is expected to contain the bulk of the virialized cluster mass and serves as a reference point for measuring different parameters of galaxy clusters in both observational and simulated galaxy catalogues.

On the other hand, to define the radius of the filaments, we have computed their radial density profile and selected halos within the regions enclosed by an overdensity $\delta + 1= 10$, since this region approximately encloses almost the entire filamentary structure \citep{Pereyra2020}. 
The values obtained for the filament radius are in quantitative agreement with \cite{Wang2024}, who studied the evolution of the filament radius in MTNG simulations and found that the typical value is constant and approximately equal to $1\ \mathrm{h^{-1}} \mathrm{Mpc}$ in the redshift range analysed in this work.
We also restrict our choice to halos with distance to the node $\text{d}_{\text{n}} > 1 \text{R}_{\text{200}}$.

In table \ref{tab:desc_samples} we show the total number and percentages of halos and subhalos for each environment and for each redshift. 
We observe that from $z=2.0$ to $z=0.0$ the fraction of galaxies in filaments decreases by about $3\%$, while in nodes it increases by about $15\%$.
The decrease (increase) of galaxies in the filaments (nodes) could be partially explained by the transport of matter through the filaments \citep{Aragon-Calvo2010,Cautun2014}, finding that the percentage of galaxies arriving at the nodes is almost five times respect to that leaving the filaments. The latter could indicate the existence of a flow of galaxies from other structures such as walls \citep{Cautun2014} or voids \citep{Lares2017}.
On the other hand, the percentage of halos increases in both structures by approximately $1\%$ for nodes and $6\%$ for filaments. The slight increase in percentage may be attributed to the limited range of redshifts considered. A larger range is necessary to observe significant changes in the fraction of halos \citep{Hahn2007}.

\section{Analysis}\label{sec:analysis}

\subsection{Evolution of HOD}

The HOD has been used to analyse the behaviour of the halo occupancy in different structures such as voids \citep{Alfaro2020}, future virialised superstructures \citep{Alfaro2021}, nodes and filaments \citep{Perez2024}. In the recent work of \cite{Perez2024} we have analysed and compared the HOD in nodes and filaments at $z=0$ and found that the distribution of halos in filaments is similar to that of the total sample, while the low mass halos in nodes show an excess of faint galaxies, which decreases as we consider brighter galaxies.

The interesting results obtained for the node and filament HODs in the local Universe motivated us to study the HOD at different redshifts for two defined cosmic structures, in order to compare the behaviour of the halo occupancy evolution. 
To measure the HOD, in this subsection we compute the mean number of galaxies per halo mass bin, $\langle \text{N}_{\text{gal}}|\text{M}_{\text{halo}}\rangle$, and estimate the errors using the jackknife technique \citep{Quenouille1949}.
The HOD measurement is computed with respect to four r-magnitude thresholds ($\text{M}_\text{r} - 5\text{log}_{10}(h) \leq -17, -18, -19\ \text{and} -20$), since these limits allow us to compare the behaviour of faint and bright galaxies in the halo \citep{Alfaro2020,Perez2024}.

\begin{figure*}
\centering
    {
        \includegraphics[width=0.85\textwidth]{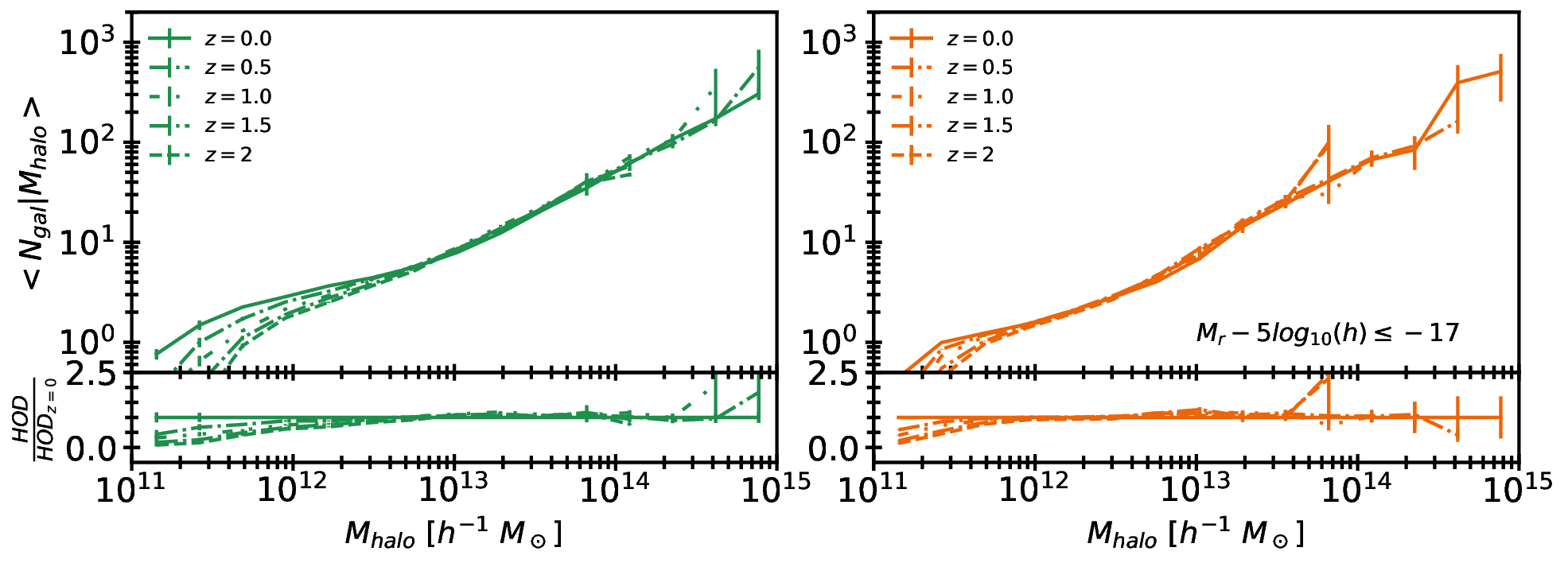}
    }
    {
        \includegraphics[width=0.85\textwidth]{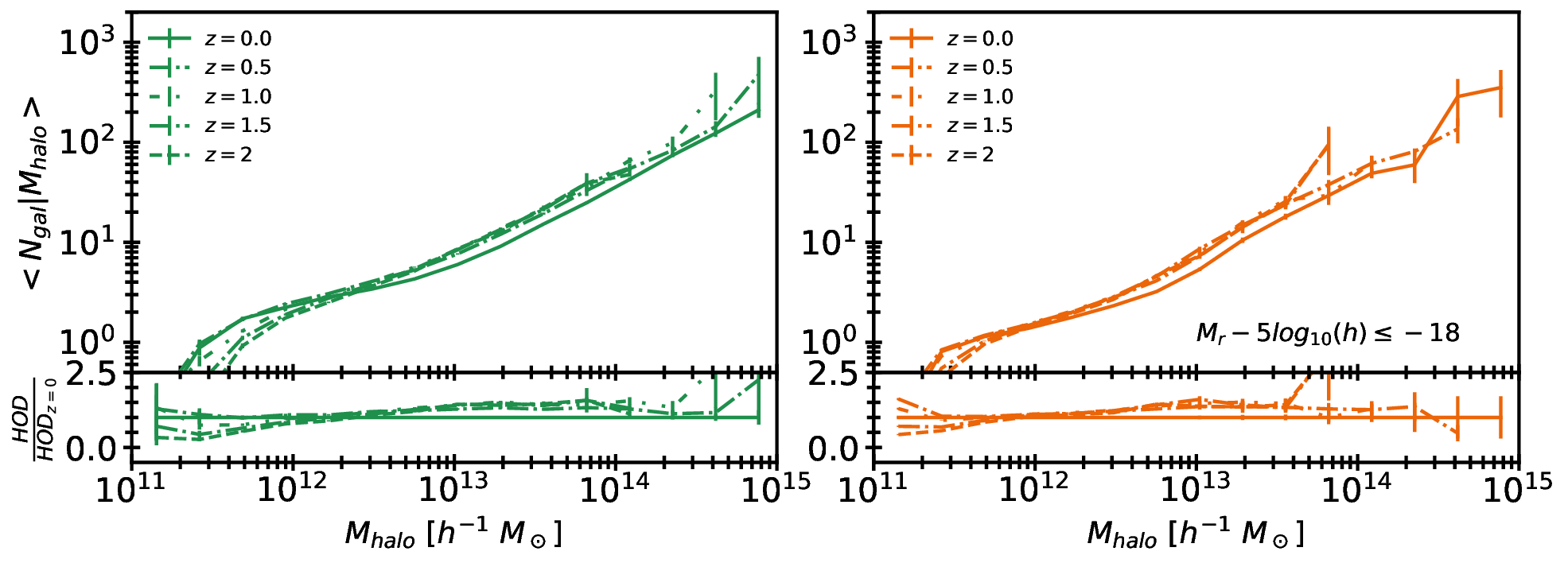}
    }  
    {
        \includegraphics[width=0.85\textwidth]{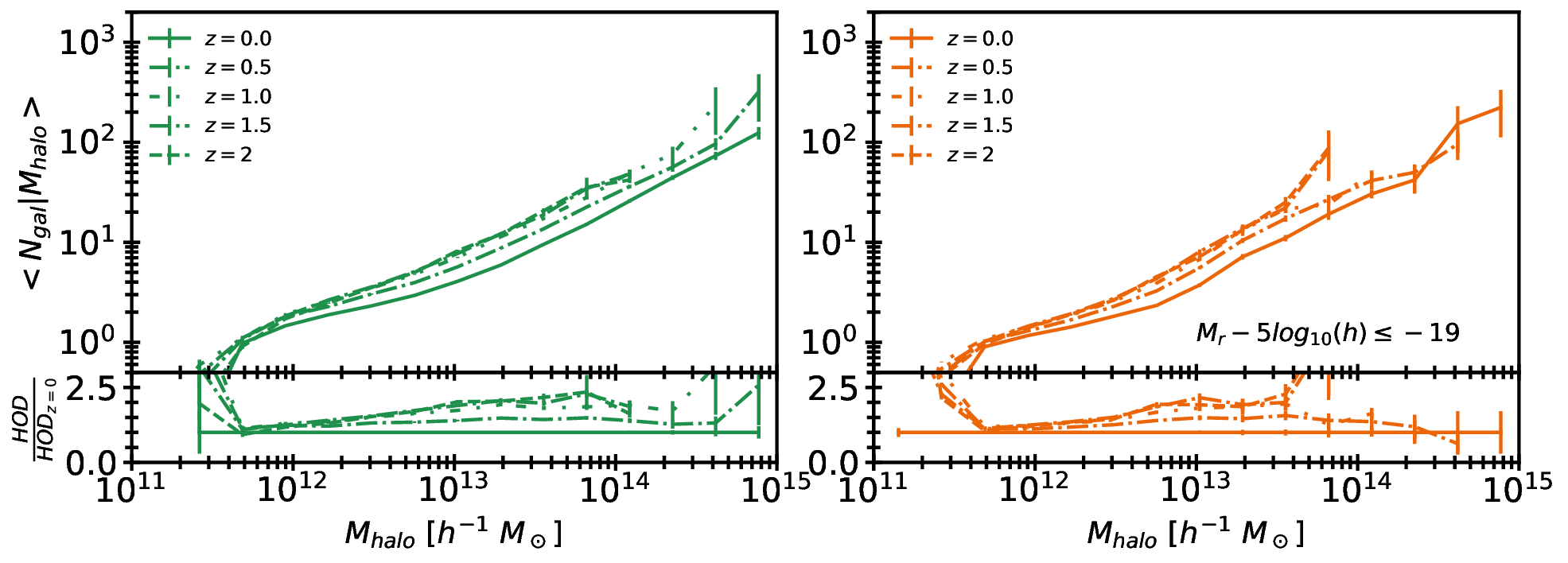}
    }  
    {
        \includegraphics[width=0.85\textwidth]{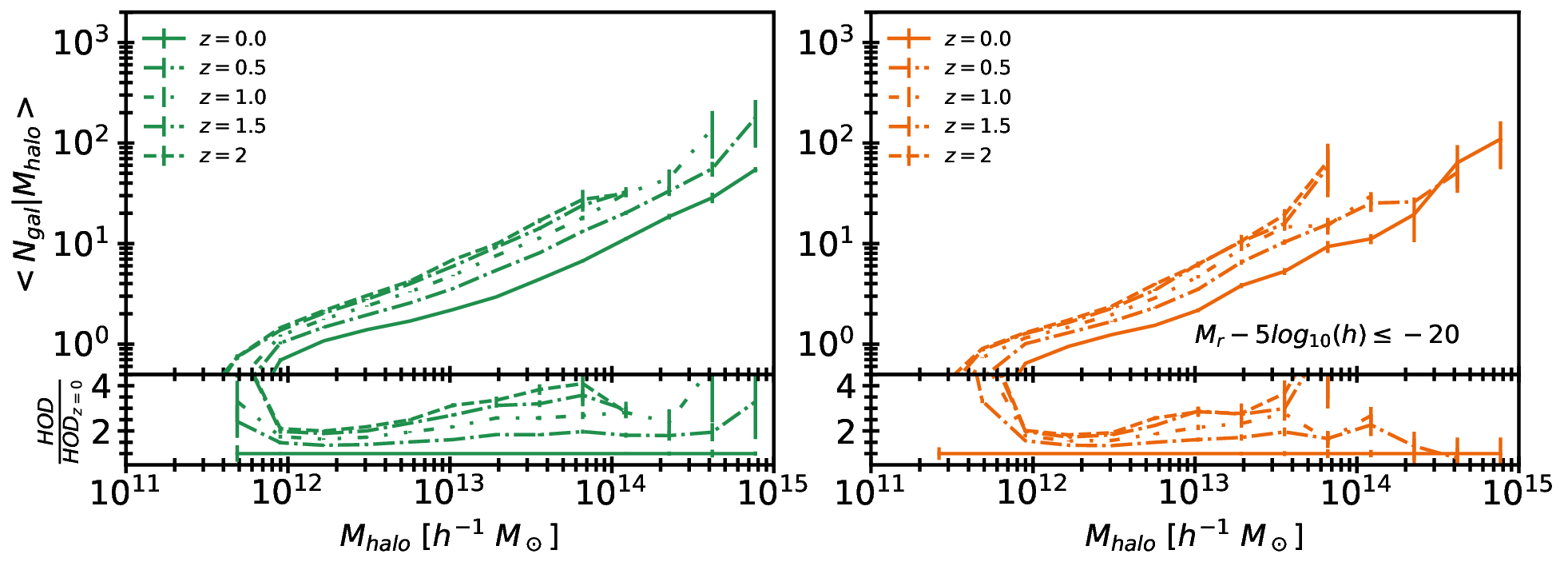}
    }        
    \caption{The panels show the HODs measured for galaxies at different redshift snapshots for the node sample (left) and for the filament sample (right), with magnitude thresholds ranging from $-17$ to $-20$ from top to bottom. The lower sub-panels show, for each environment, the ratio of the HODs at different z to the corresponding HOD in the local Universe ($z=0$)}.
    \label{fig:hod}
\end{figure*}

The top panels of Fig. \ref{fig:hod} show the HOD for the nodes and filaments at different redshifts and for all magnitude thresholds.
The lower panels show the ratio between the different HODs and their corresponding magnitude threshold sample at $z = 0$ in order to highlight their differences.
In both environments, the halo occupation decreases as we consider brighter galaxies (from top to bottom).
Furthermore, we found that the HOD curve spans an increasing range of halo masses as $z$ decreases, suggesting that massive halos are the result of evolutionary processes.
It is worth noting that halos require more initial mass to host a bright galaxies than to host faint galaxies.

Regarding the nodes, in the weaker galaxy magnitude limit ($\text{M}_\text{r} - 5\text{log}_{10}(h) \leq -17$), the HODs differ for halos with masses lower than $10^{13}\ h^{-1} \text{M}_{\odot}$, while for halos with $\text{M}_{\text{halo}}> 10^{13}\ h^{-1} \text{M}_{\odot}$ the overlap of the HODs is consistent across the entire range of redshifts, indicating that massive halos host a comparable number of faint galaxies, regardless of their redshift. 
The nodes sample exhibits a notable overabundance of faint galaxies at lower halo masses as $z$ decreases, which diminishes for bright galaxies. 
Low mass halos may have low rates of interaction and merging or may have undergone recent accretion processes, as expected at local redshift. Therefore, these areas are likely to favour the presence of faint galaxies. 
For this magnitude threshold, a similar trend is observed for halos that belong to filaments, although it is more attenuated. The HODs differ only for halos with masses smaller than $10^{12}\ h^{-1} \text{M}_{\odot}$. 
Furthermore, it is important to consider the halos within the filament sample, especially at higher redshifts, as they may be contaminated by smaller groups that have not been identified as peaks in the density field due to the persistence threshold applied in DisperSE. However, investigating the relationship between the persistence threshold and redshift is beyond the scope of our study. Quantifying the robustness of filaments is a complex task in the absence of a reference for the "true" filament axis, requiring the use of different metrics to assess the degree of overlap between different structures \citep[e.g.][]{Zakharova2023,GalarragaEspinosa2024}.

For the magnitude threshold, $\text{M}_\text{r} - 5\text{log}_{10}(h) \leq -18$, it is observed in both cosmic structures that for halos with masses lower than their respective cut-off mass, the observed trend with the redshift persists. 
Halos with masses higher than these values show the HOD overlapping over the whole range of $z$, except for $z=0.0$ where the halo occupation is lower.
For the magnitude thresholds, $\text{M}_\text{r} - 5\text{log}_{10}(h) \leq -19$, the tendency shows no difference for the low mass halos with respect to the studied redshift while the HODs become increasingly different for the high mass range.
Finally, at the brightest magnitude limit, the HODs are clearly distinguishable across the halo mass range, and the halo occupancy decreases as $z$ decreases.

\subsection{HOD parametrization}

An interesting way of quantifying the evolution is through the HOD parametrization.
\cite{Zheng2005} analysed the HOD of a smoothed particle hydrodynamics (SPH) simulation and a semi-analytic (SA) galaxy formation model, and proposed a model to describe it in terms of 5 parameters. The aim was to model the observed clustering of galaxies.
Central and satellite galaxies are considered separately because central galaxies behave differently from other galaxies in SA models and hydrodynamical simulations \citep{Kravtsov2004}. In a merger in SA models, the central galaxy of the merged halo was the most massive of the progenitor galaxies. The new central galaxy attracts the rest of the satellite galaxies through dynamic friction. Additionally, central galaxies accrete most or all of the cooling gas. 
The hydrodynamic simulations distinguish the central galaxies from the remaining galaxies by their age and mass \citep{Zheng2005}.

Since halos contain either zero or one central galaxy (usually near the centre \citep{Berlind2003}), their contribution is a step-like function, while the number of satellite galaxies has no upper limit and their contribution is a power-law-like function due to the observed proportionality between halo mass and satellite number. For more details, see \cite{Zheng2005,Zheng2007,Zehavi2011}.

The HOD for central galaxies is describe as:
\begin{equation}
 \langle N_{\rm cen}(M_{\rm h})\rangle = \frac{1}{2}\left[ 1 + {\rm erf} \left( \frac{\log M_{\rm h} - \log M_{\rm min}}{\sigma_{\log M}}  \right) \right],
\label{Eq:Cen_HOD}
\end{equation}
where $M_{\rm h}$ is the halo mass and ${\rm erf}(x)$ is the error function:
\begin{equation}
 {\rm erf}(x) = \frac{2}{\sqrt{\pi}} \int_{0}^{x} e^{-t^2} {\rm d}t.
\end{equation}
Here $M_{\rm min}$ is the characteristic minimum halo mass that can host central galaxies, i.e. the halo mass at which half the halos are occupied by a central galaxy and 
$\sigma_{\log M}$ is the characteristic width of the transition from zero to one galaxy per halo, and represents the logarithmic scatter between stellar mass and halo mass.

Then, the HOD for satellite galaxies can be represented by:
\begin{equation}
 \langle N_{\rm sat}(M_{\rm h})\rangle = \left( \frac{M_{\rm h}-M_{\rm cut}}{M^*_1}\right)^\alpha,
\label{Eq:Sat_HOD}
\end{equation}
for $M_{\rm h} > M_{\rm cut}$, where $M_{\rm cut}$ is the minimum mass of the halo hosting the satellites, $M^*_1$ is the power-law normalisation
and $\alpha$ is the slope of the power law. 
$M_1 = M_{\rm cut} + M^*_1$ is often used to measure the average halo mass for hosting a satellite galaxy \citep{Zehavi2011}.
A schematic representation of the five parameters of the HOD description can be found in Fig. 1 of \cite{Contreras2017,Contreras2023}.
Finally, the sum of the independent contributions from the central and satellite galaxies gives the halo occupation \citep{Contreras2017,Contreras2023}:
\begin{equation}
 \langle N_{\rm gal}(M_{\rm h})\rangle =  \langle N_{\rm cen}(M_{\rm h})\rangle +  \langle N_{\rm sat}(M_{\rm h})\rangle.
\end{equation}

\begin{figure*}
\centering
    {
        \includegraphics[width=0.45\textwidth]{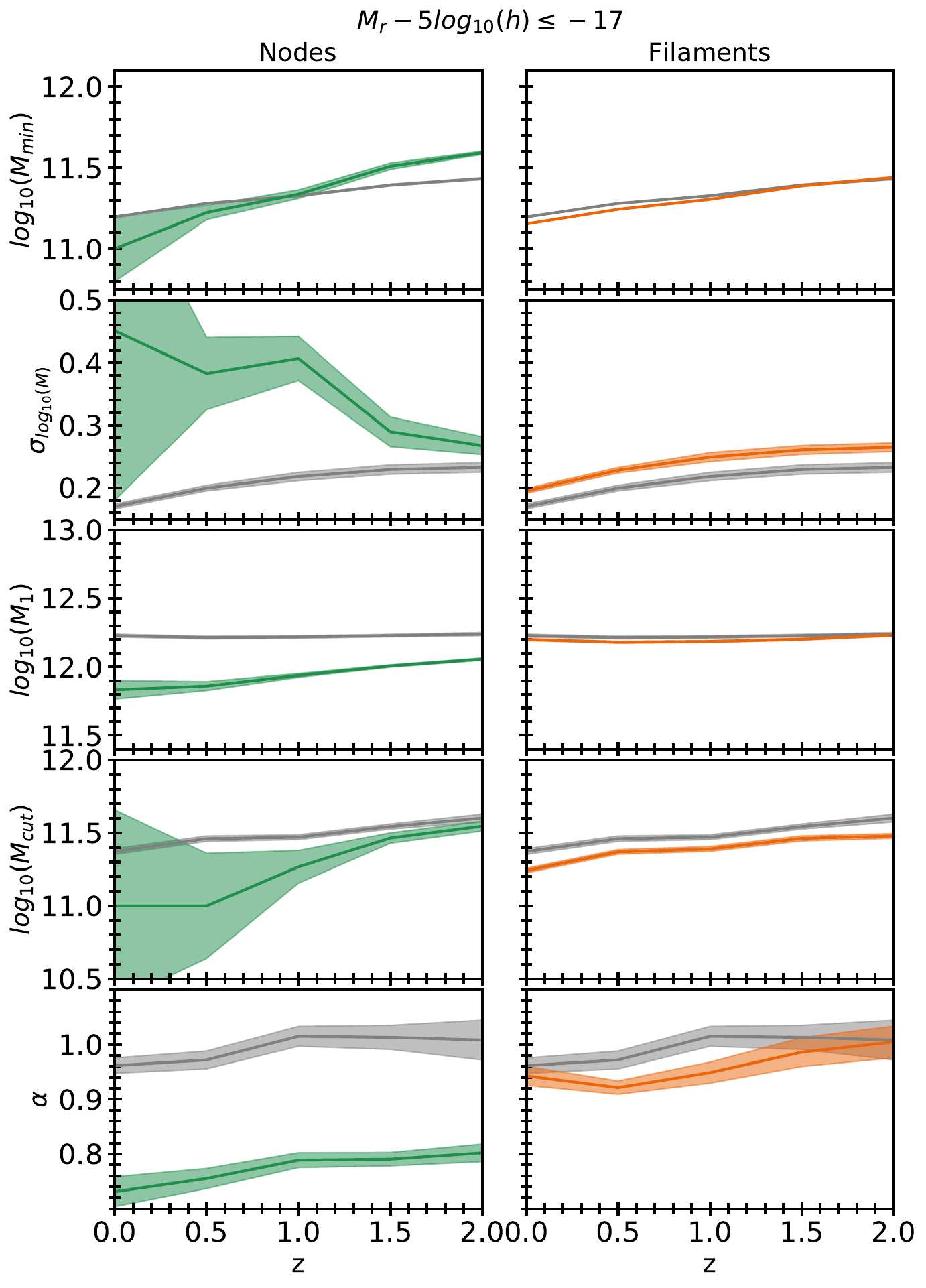}
    }
    {
        \includegraphics[width=0.45\textwidth]{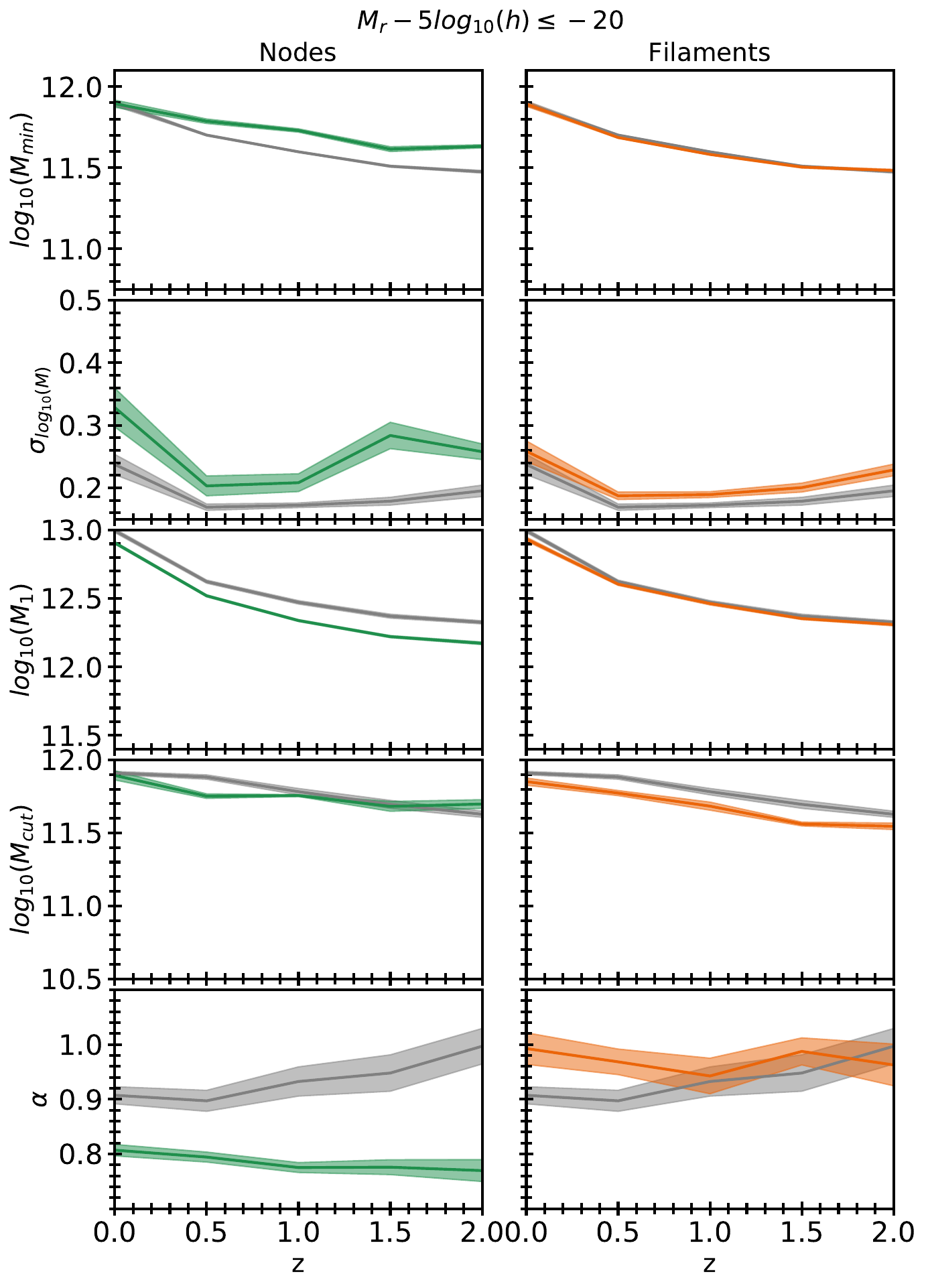}
    }      
    \caption{The evolution of the HOD parameters for total (grey), node (green) and filament (orange) samples as a function of the redshift for different magnitude
    thresholds. The shaded regions show the standard deviation from the fitted parameter value.} 
    \label{fig:parametros}
\end{figure*}

To quantify the evolution of the HOD at nodes and filaments, we fitted the five parameters of the model and detected their changes as a function of redshift and magnitude threshold.
For this purpose, the contributions from the central and satellite galaxies were fitted independently, with the jackknife error estimated per bin. 
In addition, following \cite{Contreras2023}, the errors of the fit parameters are normalised to one ($\chi^2_{\rm min} /{\rm d.o.f} = 1$) for the best parameter fit.
Figure \ref{fig:parametros} shows the variations of the parameters as a function of the redshift for nodes, filaments and the total sample for the studied magnitude thresholds.
For the sake of simplicity, only the results for the extreme absolute magnitude thresholds are presented.

The variation of $M_{\rm min}$ with redshift for both magnitude limits is shown in the upper panels.
If we look at the lower magnitude threshold, we observe for the total sample that the value of $M_{\rm min}$ increases with redshift, whereas if we look at the brighter galaxies, the trend is reversed and $M_{\rm min}$ decreases with z. 
We note that at $z=2$ the mass required for a halo to host a central galaxy does not vary significantly as a function of galaxy magnitude.
At $z=0$ there is a dependence of the mass of the halo on the magnitude of the central galaxy, indicating that the halo mass required to host a bright galaxy is higher than that required for a faint galaxy.
For all z and magnitude limits, we observe that the variation of this parameter is similar for the filament sample, suggesting that the mass at which half of the halos within filaments have a central galaxy does not vary from the mean, and hence central galaxy formation is not particularly affected when the halo belongs to a filament.
On the other hand, $M_{\rm min}$ is higher for the nodes than for the total sample at $z>1$, while at $z=0$ it is similar for both samples, regardless of the magnitude threshold.
In general, during early epochs, halos within nodes must have above-average mass to host a central galaxy. Over time, this mass requirement becomes average and even slightly lower if the central galaxy is faint. However, if the central galaxy is bright, the mass requirement is always higher than average and only coincides at $z=0$.
Therefore, the occupation of the halo by central galaxies will be affected if the halos are located in node environments.
Moreover, in all cases, we observe that at the lower limit of the magnitude threshold, $M_{\rm min}$ is lower at $z=0$ than at $z=2$. For bright galaxies, the trend is reversed, with $M_{\rm min}$ higher than the local $z$.
This result can be directly related to Fig. \ref{fig:hod}, where the tendency of the low mass HODs varies with galaxy magnitude.
For faint galaxies, the HOD at local $z$ extends to include lower-mass halos than at high $z$, while the opposite is true for bright galaxies, consistent with the detected variation in the $M_{\rm min}$ parameter.

The other parameter describing the central galaxy occupation is $\sigma_{\log M}$, which measures the smoothness of the transition from zero to one, where $\sigma_{\log M}=0$ for step transitions and $\sigma_{\log M} \not = 0$ for smooth transitions.
This parameter depends on the model used \citep{Contreras2023} and, like $M_{\rm cut}$, slightly affects the clustering of galaxies, so it is not essential \citep{Zehavi2011}. 
Furthermore, although $\sigma_{\log M}$ takes into account the physics of galaxy formation, it does not capture the dependence of stellar mass on assembly history \citep{Contreras2023}.
We observe that the total sample has higher $\sigma_{\log M}$ values for the fainter magnitude thresholds at z=2, and that they decrease slightly with time. 
For brighter galaxies, however, the trend is the opposite.
The trend for the filaments is similar to that of the total sample, although in all cases they show values above this.
The $\sigma_{\log M}$ variation for the nodes was then calculated, and significant errors in the estimation were obtained, which decrease with increasing galaxy magnitude.
Note, however, that the values obtained for the full range of redshift and for all magnitude limits are always higher than those for the total sample.
This behaviour of $\sigma_{\log M}$ is consistent with Fig. \ref{fig:hod}, where the smoothness of the transition from zero to a galaxy can be observed (for more details see Fig. 1 of \cite{Contreras2017,Contreras2023}).
These results suggest that the stellar mass - halo mass dispersion is higher for halos belonging to nodes or filaments,
indicating that the formation of central galaxies of a given stellar mass is less closely related to a specific range of halo masses than expected.

\begin{figure}
\centering
    {
        \includegraphics[width=0.45\textwidth]{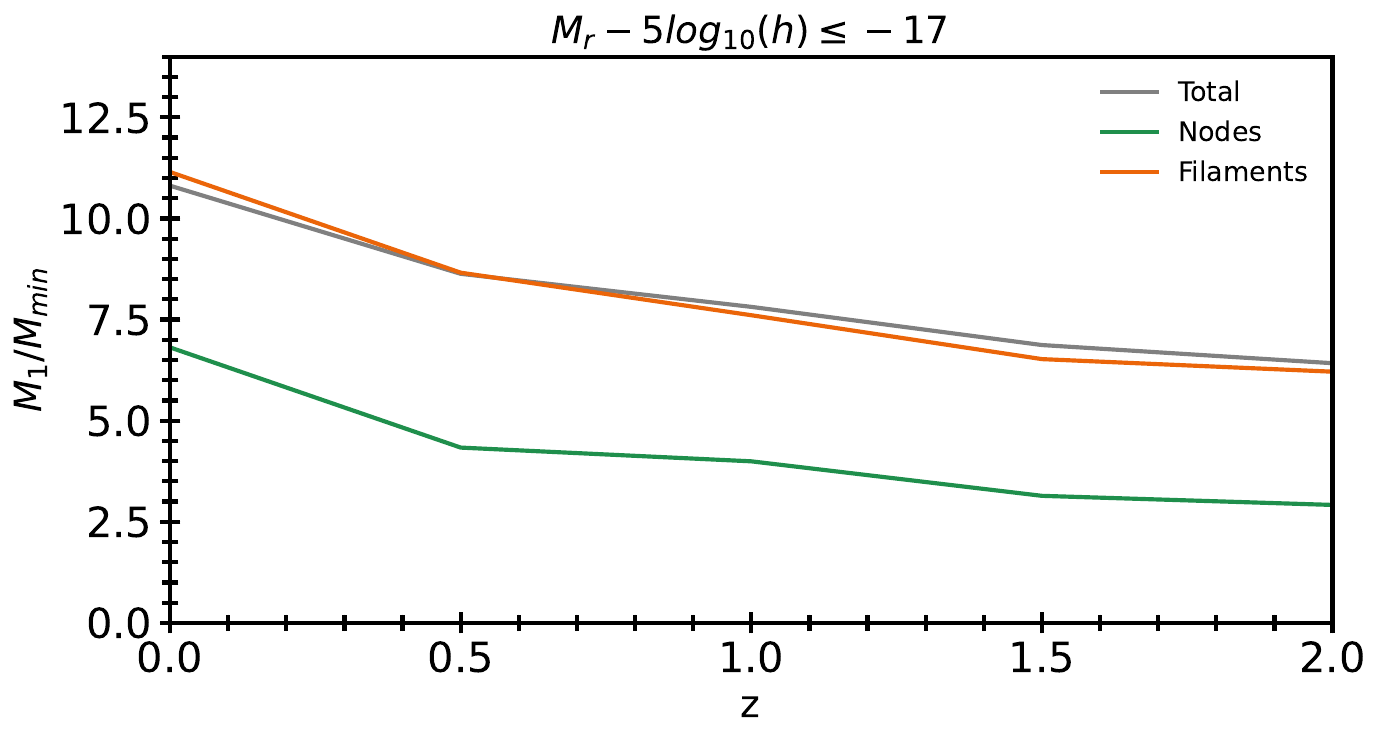}
    }
    {
        \includegraphics[width=0.45\textwidth]{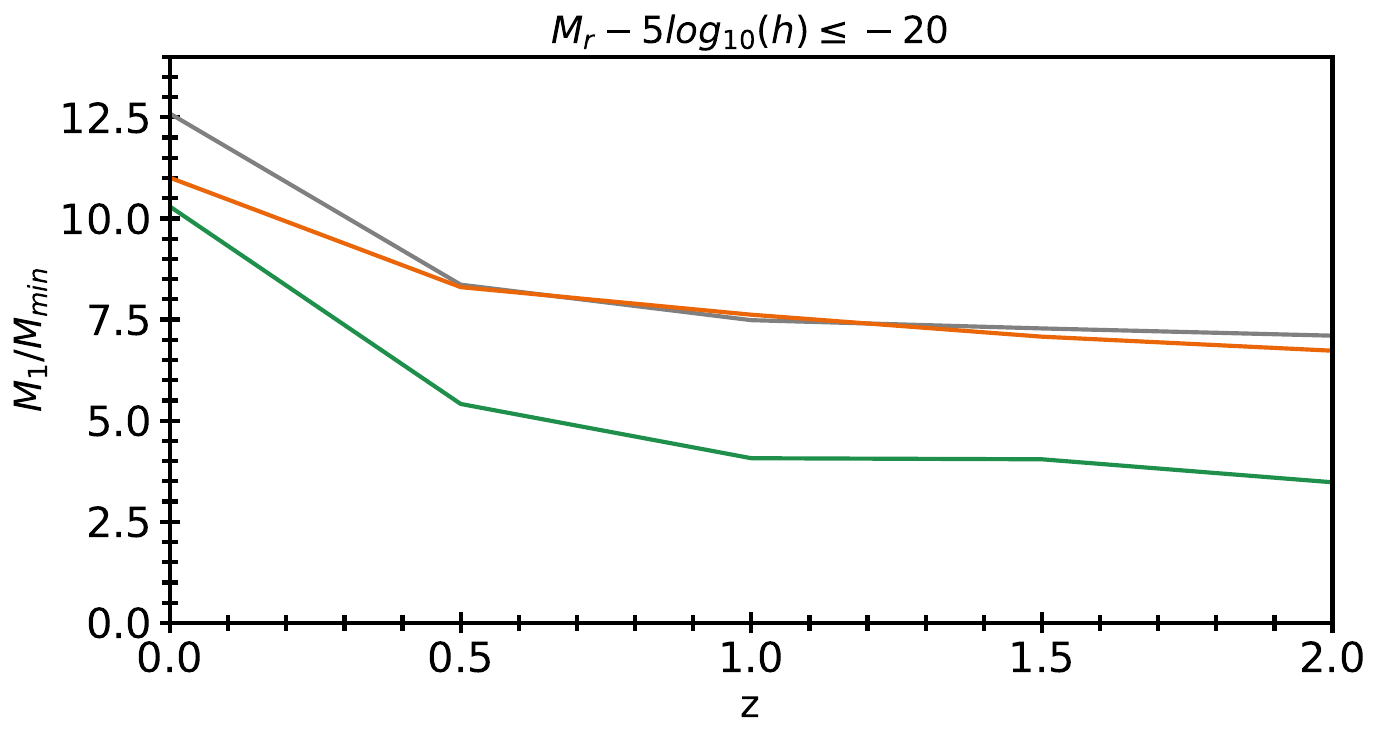}
    }      
    \caption{Redshift evolution of the ratio of
     the two characteristic HOD paramenters ($M_{1}$ and $M_{\rm min}$) for total, node  and filament samples with respect to the redshift for different magnitude thresholds.}
    \label{fig:cociente}
\end{figure}

On the other hand, three parameters are used to determine the contribution of satellite galaxies: two parameters related to the masses and the slope of the curve, described below.
$M_1$ quantifies the halo mass at which, on average, a halo will have one satellite galaxy.
$M_1$ shows no significant variation with redshift for the weakest magnitude threshold and for all three samples.
While the values for the total sample and the filament sample are similar, $M_1$ is lower for the node sample over the whole redshift range.
However, when considering brighter galaxies, $M_1$ increases from $z=2$ to $z=0$, in agreement with \cite{Contreras2023}, with the node sample also having lower values independently of the redshifts.
Wherever the halo is located, it needs more mass at $z=0$ than at $z=2$ to host a bright satellite galaxy.
Both the total and filament samples show similar behaviour of $M_1$, indicating that the mean mass at which halos in filaments have at least one satellite galaxy is not different from the average.
Nodes require less mass than average to host a satellite galaxy. 
This may be due to the continuous accretion of material, so they do not need extra mass to get a satellite galaxy.
This suggests that dense environments favour the presence of satellite galaxies in low mass halos and thus lead to an abundance of these galaxies in low mass halos \citep{Van2016}.
%

$M_{\rm cut}$ indicates the minimum halo mass required to host a satellite and is the other mass involved in the HOD description for satellite galaxies.
For the whole sample and considering the lower limit of magnitude, $M_{\rm cut}$ increases slightly with redshift, in agreement with \cite{Contreras2023}.
As the galaxies become brighter, the trend reverses, with $M_{\rm cut}$ being larger at local $z$ than for high values of $z$.
This means that at $z=0$ the minimum mass to host a satellite galaxy depends on the magnitude of the galaxy in question, being lower if it is a faint galaxy. 
Faint satellite galaxies are then more likely to form in low mass halos, while bright ones are more likely to be hosted by massive halos.
Furthermore, at $z=2$ the minimum mass is independent of magnitude.
The trends for halos belonging to filaments and nodes are similar to those for the total sample, but with lower values, indicating that halos in these regions have satellite galaxies, even if they are less massive than the average.
This means that the minimum halo mass required to host a satellite galaxy is significantly influenced by these environments.

\cite{Contreras2017} found that the evolution of the power-law slope has a small effect on the evolution of the HOD at low redshift.
In general, this parameter is close to 1, although for bright galaxies ($M_r < -22$) it is much larger, with significant errors \citep{Zehavi2011}.
The observed trends show lower values at $z=0$, which start to increase slightly with increasing z.
We also find slope values close to $\alpha=1$,
for the total and filament samples, while the node sample has values around $\alpha=0.8$. 
This suggests that if the halo is within a filament, the number of satellite galaxies it hosts increases linearly with the mass of the halo, whereas if the halo belongs to a nodes, it increases more slowly, since the formation and distribution of satellite galaxies is affected in a non-trivial way by the properties of these dense environments.
As the difference in the slopes of the two samples is not significant, it is barely visible in the Fig. \ref{fig:hod}.

The Fig. \ref{fig:cociente} shows, for the extreme absolute magnitude thresholds, the evolution of $M_{1}/M_{\rm min}$, which is usually used to characterise the relationship between the halo mass required to host a central galaxy and to populate the halo with a satellite galaxy, indicating how efficient a halo is at hosting satellite galaxies relative to the presence of a central galaxy \citep{Zehavi2011,Coupon2012}.
We observed a similar behaviour for both magnitude thresholds, with the ratio being higher at $z=0$ and decreasing at high z.
This suggests that central galaxies are currently populating halos over a wider range of halo masses before being occupied by satellite galaxies than at earlier times.
Furthermore, these high values of $M_{1}/M_{\rm min}$ are reflected in the plateau feature in the HODs of Fig. \ref{fig:hod}.
The $M_{1}/M_{\rm min}$ values for the total and filament samples are close, while for the node sample the rate is lower across the redshift range.
This means that halos within nodes are populated by satellite galaxies much earlier than if the halo were in a filament.
The trends observed are consistent with \cite{Contreras2023} and with the tendencies of $M_{\rm cut}$.

It is important to highlight that in \cite{Contreras2023}, the analysis is performed on stellar-mass selected galaxy samples with fixed number densities.
Our results emphasise their finding, which exposes that this ratio does not contain key information about galaxy formation physics.
This reinforces that the evolution of the HOD may be independent of galaxy formation physics or the underlying cosmological framework. 

\subsubsection{HOD parameters: direct measurement}

\begin{figure*}
\centering
    {
        \includegraphics[width=0.8\textwidth]{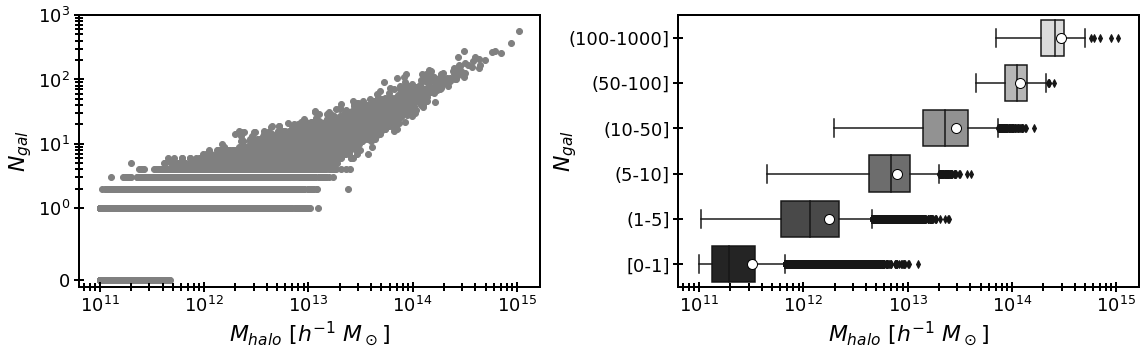}
    }
    {
        \includegraphics[width=0.8\textwidth]{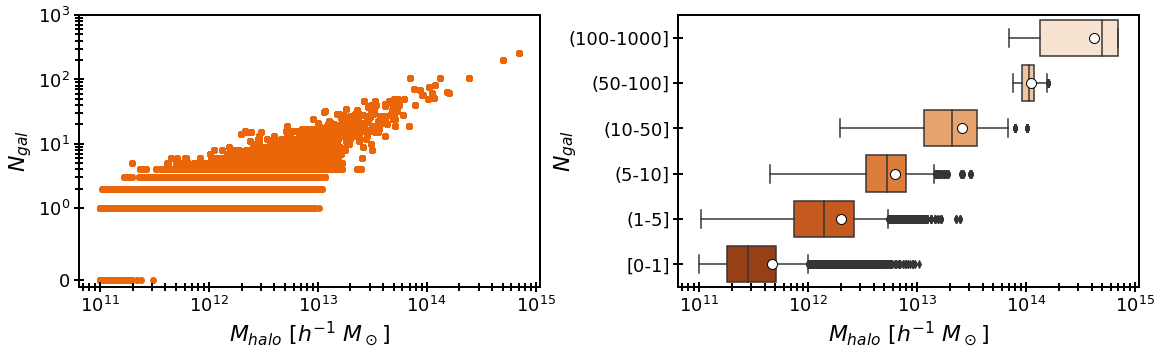}
    }   
    {
        \includegraphics[width=0.8\textwidth]{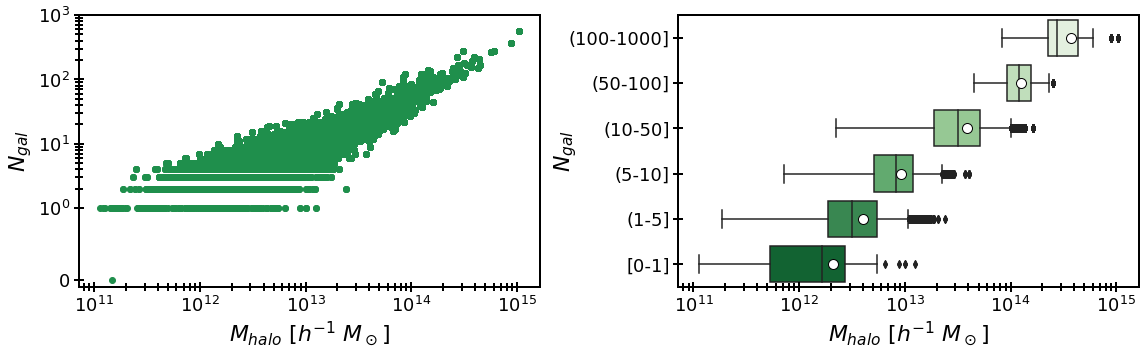}
    }       
    \caption{Galaxies with $\text{M}_\text{r} - 5\text{log}_{10}(h) \leq -17$ at $z=0$ from total, node and filament samples from top to bottom.
    Left panels: Scatter plots of raw simulation data before plotting HOD. 
    Right panels: Box plots for six ranges of galaxy numbers.
    The line within the box denotes the median of the dataset, while the white point indicates the mean value. 
    The box extends from the lower quartile (25th percentile) to the upper quartile (75th percentile).
    The individual dots beyond these whiskers are considered outliers.}
    \label{fig:scatter-boxplot}
\end{figure*}

The methodology described above for parametrizing the HOD is based on a model fit derived directly from the HODs illustrated in Fig. \ref{fig:hod}. However, some of the parameters can also be measured directly from the dataset.
The data from the simulation, illustrated in the left panels of Fig.  \ref{fig:scatter-boxplot}, which depicts the relationship between the number of galaxies and halo mass at $z=0$, exhibits a noisy scatter plot. This noisy dispersion is observed in all redshift snapshots. For simplicity, only the results for $z=0$ are presented.
To derive the HOD function, we perform binning on the halo mass, which enables us to compute a more accurate and manageable representation of the data distribution along this axis.
Simultaneously, we calculate the mean number of galaxies for each bin on the y-axis.     
Then, comparing the parameters obtained from the original data will reveal discrepancies, which are primarily influenced by the number of bins used and the intrinsic properties of the original data.

To quantify the dispersion of the data, we show box plots for six ranges of galaxy numbers in the right panels of Fig. \ref{fig:scatter-boxplot}.
Note that there are significant outliers that may not be represented in the measured HOD.
The resolution and smoothness of the HOD plot are influenced by the number of bins used in the calculation. 
A higher number of bins provides a more detailed representation but is more susceptible to data noise. 
Conversely, using fewer bins yields a more generalised view, which may result in a loss of information and fail to capture the complex characteristics of the original data set.

\begin{figure}
    \centering
    \includegraphics[width=0.99\linewidth]{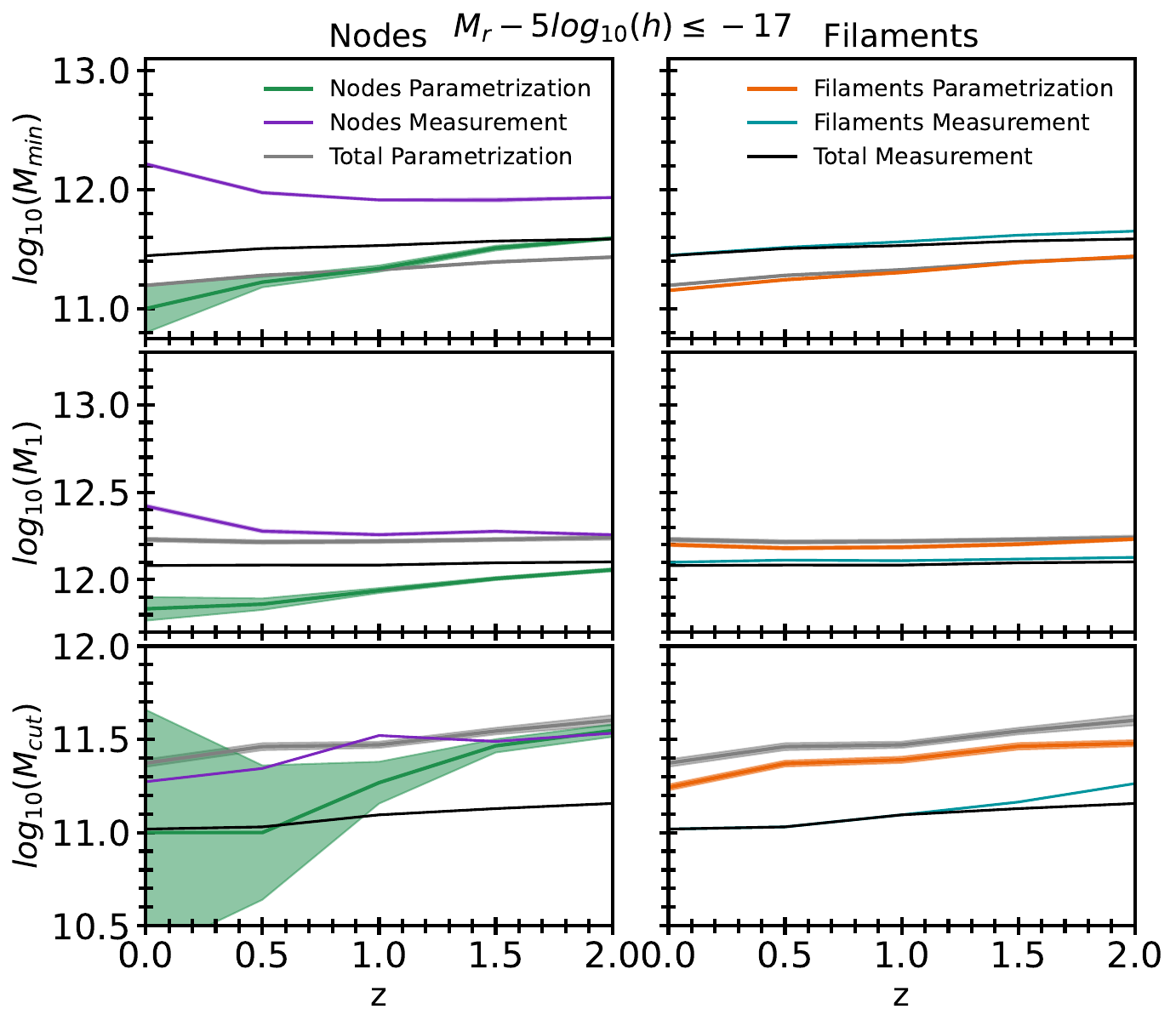}
    \caption{Evolution of the mass parameters directly measured from the simulation, and derived from the parametrization, for the total, node, and filament samples as a function of redshift for galaxies with $\text{M}_\text{r} - 5\text{log}_{10}(h) \leq -17$.}
    \label{fig:comparación-masas}
\end{figure}

To illustrate the possible discrepancies, we have directly measured $M_{min}$, $M_1$ and $M_{cut}$ from the simulation for galaxies with $\text{M}_\text{r} - 5\text{log}_{10}(h) \leq -17$ at all z, as shown in Fig. \ref{fig:comparación-masas}.
In order to ascertain the values of the aforementioned parameters, we have adopted the definitions utilised in our parametrization. In this context, $M_{min}$ represents the halo mass at which half of the halos are occupied by a central galaxy, $M_{cut}$ denotes the minimum mass of the halo hosting the satellites, and $M_1$ signifies the average halo mass for hosting a satellite galaxy. Moreover, these values can also be identified in the box plots presented in Fig. \ref{fig:scatter-boxplot}.

In all cases, the parameters measured for the filament sample are in alignment with those observed in the total sample. Moreover, the $M_{min}$ and $M_1$ parameters, both estimated and directly measured, demonstrate a comparable trend for the filament sample. However, a notable discrepancy exists between the two estimated values of $M_{cut}$ for the filament sample. Additionally, the three parameters obtained for the node sample exhibit a significant deviation from the observed trends in the total sample.

To gain insight into these discrepancies, we can compare the results displayed at $z=0$ of Fig. \ref{fig:comparación-masas} with the box plot of Fig. \ref{fig:scatter-boxplot}. The discrepancy between the $M_{min}$ values derived from the parametrization fit and those measured for faint galaxies in nodes can be attributed to the broad range of halo masses that can host a single galaxy ($[0,1]$), which results in a higher median value for the distribution.
The box plots for the total and filament samples are comparable and exhibit lower values than the node sample, as evidenced by the parameter comparisons.

To determine $M_{cut}$, we examine the left whisker of box plots in range $(1,5]$. It is observed that both the filament and total samples have an $M_{cut}$ value of $10^{11}\ h^{-1} \text{M}_{\odot}$, which represents our minimal halo mass according to the criteria employed in the selection of the samples (see Sect. 2.3).
This indicates that, within the simulation, halos host satellite galaxies at lower halo masses, and these halos are part of filaments. 
This observation aligns with the trend that filaments are less massive than nodes, which necessitates higher halo masses to host satellites.
These findings are evident in Fig. \ref{fig:comparación-masas}.
The bias in the total sample and filaments is not reflected in the parameters derived from the parametrization.
This is because the fit is based on a HOD model, as shown in Fig. \ref{fig:hod}.
Finally, the approximate value of $M_1$ can be inferred from the white point, which represents the average value, in box plots within the $(1,5]$ range, as it contains halos that have between one and four satellite galaxies.
The $M_1$ values measured and parametrized for the filament and total samples are similar, whereas the node sample shows significant differences.

It is important to note that the value of the HOD mass parameters depends on the method used for their measurement. 
Consequently, the resulting values may exhibit minor or substantial discrepancies when obtained through parametrization or direct measurement, respectively.
In order to achieve a more accurate characterisation of the results, given the high dispersion of the data affecting the direct measurements, it is considered that the HOD parametrization taken from the fit curve is the most suitable for the purposes of this work.
Moreover, this methodology allow us compare our results with different works in bibliography \citep[e.g.][]{Zheng2005,Contreras2023}.

\subsection{Evolution of galaxy properties}
\subsubsection{Stellar-Halo mass relation}

\begin{figure*}
\centering
    {
        \includegraphics[width=0.75\textwidth]{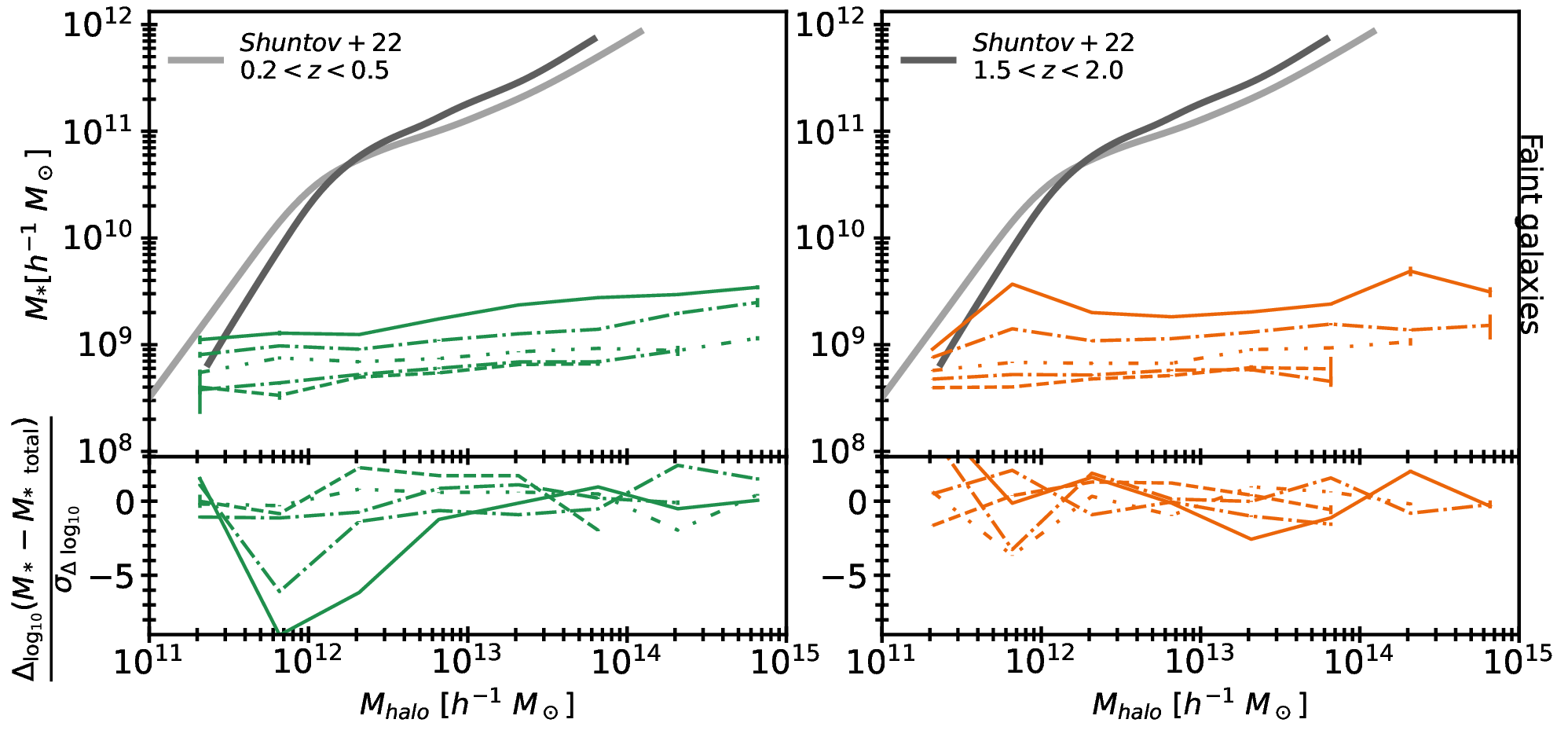}
    }
    {
        \includegraphics[width=0.75\textwidth]{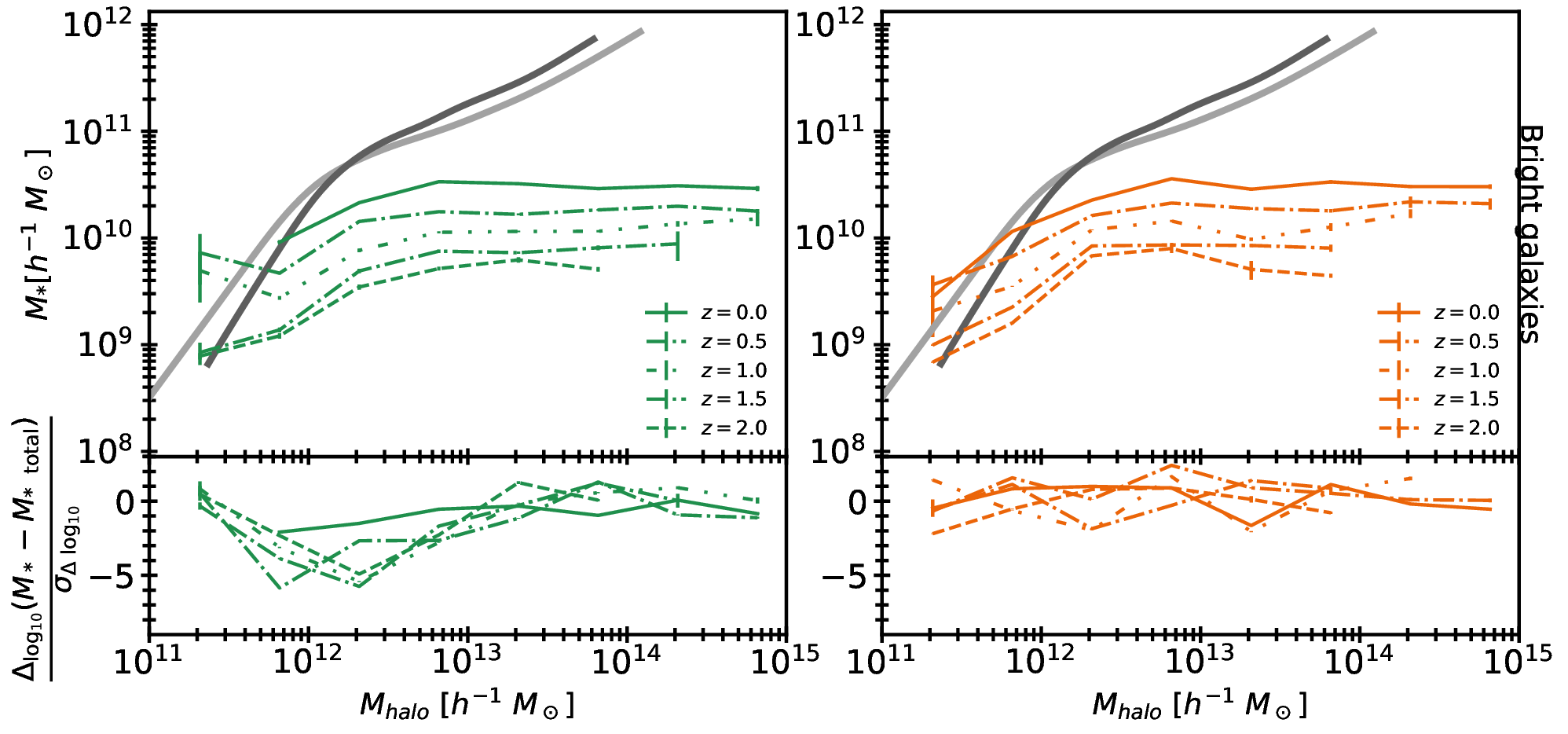}
    }       
    \caption{The stellar-to-halo mass relation (SHMR) for faint (top) and bright (bottom) galaxies belonging to nodes (left) and filaments (right). The subpanels show the logarithmic relative difference between SHMR and the total sample at the corresponding z divided by their associated error.} 
    \label{fig:shmr}
\end{figure*}

To study how galaxy properties evolve in a cosmological context, we have estimated the stellar-halo mass relation (SHMR, \cite{Wechsler2018,Scholz-Diaz2022}).
This tool links galaxies and their host halos, allowing it to be tested against models of galaxy evolution \citep{Martin-Navarro2022}.
By linking galaxies to their parent halos, the physical processes responsible for galaxy growth, which are expected to depend primarily on halo mass, can be more clearly identified and constrained \citep{Behroozi2010}. 
Furthermore, the SHMR is indicative of the efficacy of the stellar mass assembly of a galaxy, calculated over the lifetime of the halo \citep{Somer2015}. This encompasses the contributions of both the central and satellite galaxies to the total stellar mass. The majority of the contribution to the total stellar mass content in $M_{halo} \la 10^{12} h^{-1} M_{\odot}$  halos comes from central galaxies, while in $M_{halo} \ga 10^{12} h^{-1} M_{\odot}$ halos it comes from satellites \citep{Leaut2012,Shuntov2022}.
\cite{Ishikawa2017} refers to this halo mass as the characteristic dark halo mass for galaxy formation.
Here, the feedback processes affecting star formation efficiency overlap. In lower mass halos, the supernova feedback is more effective at reheating and ejecting gas, while in massive halos the AGN feedback is more effective \citep{Moster2010}.
Furthermore, in halos with masses below the characteristic mass, stellar mass growth in galaxies is mainly driven by star formation, whereas in halos with higher masses, mergers are responsible \citep{Shuntov2022}.

Figure \ref{fig:shmr} shows the SHMR from $z=0$ to $z=2$, for node and filament samples.
Furthermore, two SHMR for $0.2<z<0.5$ and $1.5<z<2.0$ are overplotted using observational data extracted from \cite{Shuntov2022}.
In order to identify potential differences, the lower panels show the logarithmic relative difference between SHMR and the total sample at the corresponding z divided by their associated error. 
The results presented in Fig. \ref{fig:hod} clearly demonstrated a magnitude dependence. To gain further insight, the properties were evaluated by discriminating between galaxies on the basis of their magnitude. The faint galaxy category was defined as those with $-20 < \text{M}_\text{r} - 5\text{log}_{10}(h) \leq -17$, while the bright galaxy category was defined as those with $\text{M}_\text{r} - 5\text{log}_{10}(h) \leq -20$.
Our analysis shows that the stellar mass content increases with time regardless of the galaxy luminosity and cosmic structure considered. Furthermore, this increase is progressive. 
At high redshifts, the difference in stellar mass between adjacent redshifts is relatively minor. In contrast, at local redshifts, the increase in stellar mass is more pronounced.
The outcomes for low mass halos agree with those presented by \cite{Moster2010}, although there are discrepancies for massive halos.

The upper panels of Fig. \ref{fig:shmr} show the SHMR for faint galaxies.
As the mass of the host halo increases, the masses of the faint galaxies associated with the nodes increase monotonically, whereas for the filament sample this dependence is noisy, especially for low mass halos at $z\leq0.5$, where a peak in star formation is observed at halo masses of approximately $10^{12} h^{-1} M_{\odot}$.
Furthermore, as illustrated in Fig. \ref{fig:hod}, the formation of massive halos in filament sample is not yet evident at $z=2$. 
The lower sub-panels indicate that at all $z$, galaxies in filaments have a similar stellar mass content to the total sample 
in the range of deviation.
Conversely, galaxies in halos with masses below approximately $10^{12.5} h^{-1} M_{\odot}$, which are associated with nodes, have lower stellar masses than the total sample at $z = 0.0$ and $z=0.5$.
The observed effect may be attributed to the fact that as galaxies become less massive, the scatter in the relation increases \citep{Engler2021}.
Meanwhile, at higher redshifts the curves show no evident trend in the dispersion behaviour.

On the other hand, the trends observed in bright galaxies (lower panels of Fig.\ref{fig:shmr}) are comparable between the two cosmic structures.
For all redshifts, the stellar mass content increases with the halo mass. 
While this increase is significant for galaxies belonging to halos with masses below approximately $10^{12.5} h^{-1} M_{\odot}$, it is more moderate for galaxies in massive halos.
Moreover, according to Fig. \ref{fig:hod}, bright galaxies at $z=0$ occupy halos with higher masses than other $z$.
As is the case for faint galaxies, bright galaxies in filaments have a similar stellar mass content to the total sample at all $z$ values.
Furthermore, nodes impact the stellar mass content in bright galaxies belonging to halos with $M_{\text{halo}} \la 10^{12.5} h^{-1} M_{\odot}$. 
This effect is less pronounced than in faint galaxies and, conversely, more significant at high redshift and negligible at local z.

A comparison between the SHMR obtained from observational data and that derived from the \textsc{TNG300} simulation consistently shows higher values for the observational data.
This discrepancy may be attributed to differences in the methodologies employed for estimating galaxy stellar masses and galaxy magnitudes, as well as the star formation models. 
We used a stellar mass definition based on the surface brightness profile, which can lead to significant differences from other methods. 
For example \cite{Pillepich2018_b} calculate the SHMR using a fixed aperture radius for their measurements (see Figure 11 of their paper). This method involves assessing the stellar mass within a predetermined physical radius around each galaxy, which might not fully account for variations in galaxy size distributions. 
They found a strongly peaked SMHR function at $z=0$ in $10^{12} h^{-1} M_{\odot}$, with a rapid decline in both lower and higher mass ranges. 
Conversely, at high redshifts, they only identified the peak and the drop at high masses.
The observed peak in star formation is in accordance with our findings. Additionally, the rate of decline at higher masses is comparable, despite the fact that they reach higher mass values.

On the other hand, \cite{Shuntov2022} compare the SHMR for the \textsc{COSMOS2020} survey \citep{Weaver2022} with three hydrodynamical simulations: \textsc{Horizon-AGN} \citep{Dubois2014,Kaviraj2017}, \textsc{Illustris TNG100} and \textsc{EAGLE} \citep{Crain2015,Schaye2015}. 
They found a good agreement in the low mass halo region, but above the peak, a significant discrepancy was observed in the SHMR.
Additionally, they quantified the satellite fraction, finding it to be higher in the simulations. 
Given the significant contribution of the satellite component to the total SHMR in this region, they concluded that the feedback mechanisms in simulations are less effective in quenching the mass assembly of satellites, resulting in a delayed quenching process.
In addition, the authors have identified a number of potential sources of error in their measurements.
Firstly, the comparison between simulations and a survey with a relatively small cover volume could result in a biased outcome. 
Secondly, uncertainties may arise from the estimation of redshift and stellar masses, clustering properties, the radial distribution of dark matter and galaxies within halos, and halo mass definitions, among others.
These potential sources of error could result in different SHMR measurements.

In this context, our results differ from those reported by \cite{Shuntov2022}. 
The authors selected galaxies with a brightness greater than [3.6]lim = 26 in the different z-slices.
However, we do not have access to this magnitude band and have divided the total sample into bright and faint galaxies with respect to the r band.
Despite that, their selection of brighter galaxies results in an SHMR range comparable to that measured for our sample of bright galaxies.
As outlined by \cite{Shuntov2022}, several factors, including ram-pressure, tidal stripping, harassment by other satellites, and a lack of pre-processing by stellar radiative feedback, could impact the stellar mass content across the entire redshift range.

Furthermore, it is expected that low-mass halos will host galaxies that are more massive at low redshift than at high redshift. In contrast, the opposite trend is observed in massive halos, which host more massive galaxies at high redshift and less massive galaxies at low redshift \citep{Moster2010}. 
Their results suggest that the growth of low mass halos is slower than that of the central galaxies they host, while large halos accrete dark matter at a rate exceeding the stellar mass growth of their central galaxies.
Our findings align with the low mass halo hypothesis, but not with the reversal observed in the case of massive halos.
This suggests that the feedback mechanisms regulating star formation in low mass halos in the simulations align with observations. 
However, the environmental quenching of satellites appears less efficient in the simulations.

\cite{Darvish2017, Yang2022} suggest that the SHMR may be influenced by the larger-scale environment, potentially leading to variations between the SHMR observed in filaments and that of the Universe as a whole. 
Nevertheless, the observed trends are largely similar, and the considerable uncertainties preclude a definitive conclusion.
Despite the fact that the classification methods used to ascertain the cosmic web structures may result in variations in the structure properties \citep{Zhu2024}, our findings are in alignment with those of \cite{Darvish2017}. 
This outcome confirms that both structures exert a comparable influence on the SHMR, with only minor discrepancies observed at local z levels regarding faint galaxies.

\subsubsection{Galaxy colours}
It is well established that a significant property that evolves over cosmic time is the colour of galaxies. Variations in galaxy colour are indicative of changes in mean age, metallicity and dust content \citep{Bell2000}. In this context, 
red galaxies are predominantly massive early-type galaxies, populated by old stars with negligible star formation rates, meanwhile, blue galaxies are star-forming and late-type objects \citep{Strateva2001,Bernardi2003,Faber2007,Schawinski2014}.
The galaxy colours show a bimodal distribution at local redshift, being either red or blue, indicating that the galaxies are populated by two dominant stellar populations. 
There are also a small number of galaxies in the green valley, which would be a mixture of blue and red galaxies \citep{Schawinski2014,Bravo2022}.
The colour bimodality is also observed at high redshifts, where the blue component dominates while the red population increases at local redshift \citep{Wolf2003,Bell2004}.

In this sense, \cite{Nelson2018} studied the galaxy flux in the $g - r$ colour-mass plane for TNG300-1 simulations from $z=0.3$ to $z=0.0$. 
They found a demarcation line in this plane at $g - r \backsimeq 0.35$ for galaxies with stellar masses below $10^{11} M_{\odot}$. This line bounds the mean change in galaxy colour with time.
The colours of galaxies with $g - r < 0.35$ become bluer with increasing stellar mass, whereas the opposite is true for galaxies with $g - r > 0.35$.
This limit is observed to shift to $g-r = 0.7$ for more massive galaxies.
The population with colour $g-r > 0.7$ reddens, but more slowly, while those with colours below this value turn blue due to episodes of rejuvenated star formation, but these are rare events as there are no massive blue galaxies.

Furthermore, within the context of cosmic structures, there is a correlation between galaxy colour and morphology trends and density.
Morphological segregation can be observed within clusters, with elliptical galaxies typically occupying the inner regions and spirals the outer regions \citep{Dressler1980,Coenda2009}.
Moreover, \cite{Kraljic2018,Laigle2018} observed a segregation of type and colour within the filaments. Specifically, they found that red and quiescent galaxies are more likely than their counterparts to inhabit regions close to the axis of the filament. 
This segregation is clearly observed for galaxies belonging to low mass halos, while the tendency is weak for galaxies within massive halos \citep{Perez2024}.

\begin{figure*}
    \centering
    {
        \includegraphics[width=0.9\textwidth]{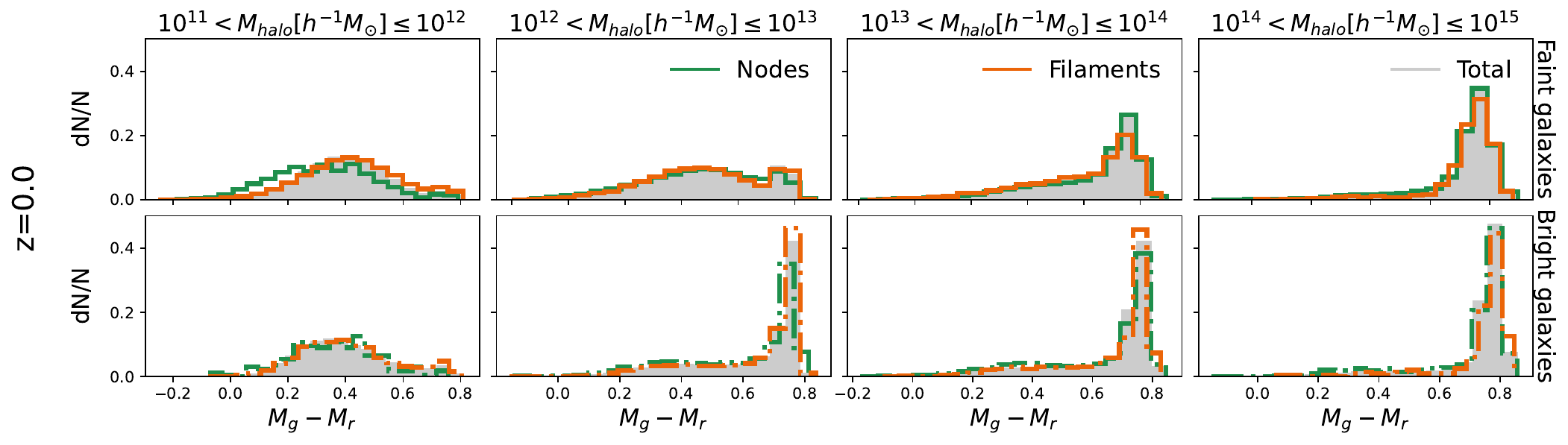}
    }
    {
        \includegraphics[width=0.9\textwidth]{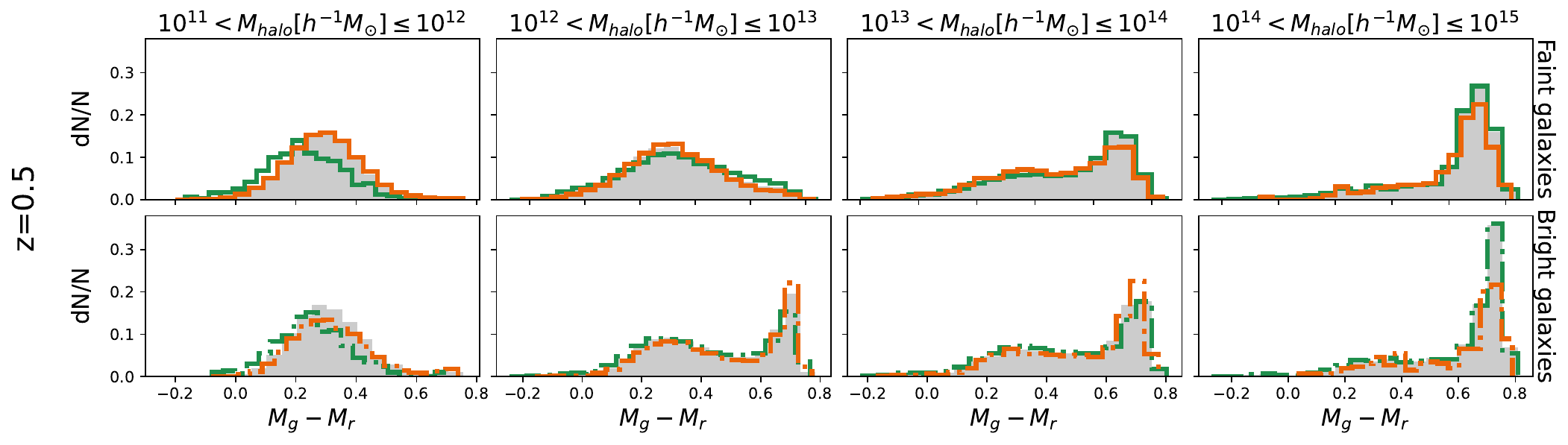}
    }
    {
        \includegraphics[width=0.9\textwidth]{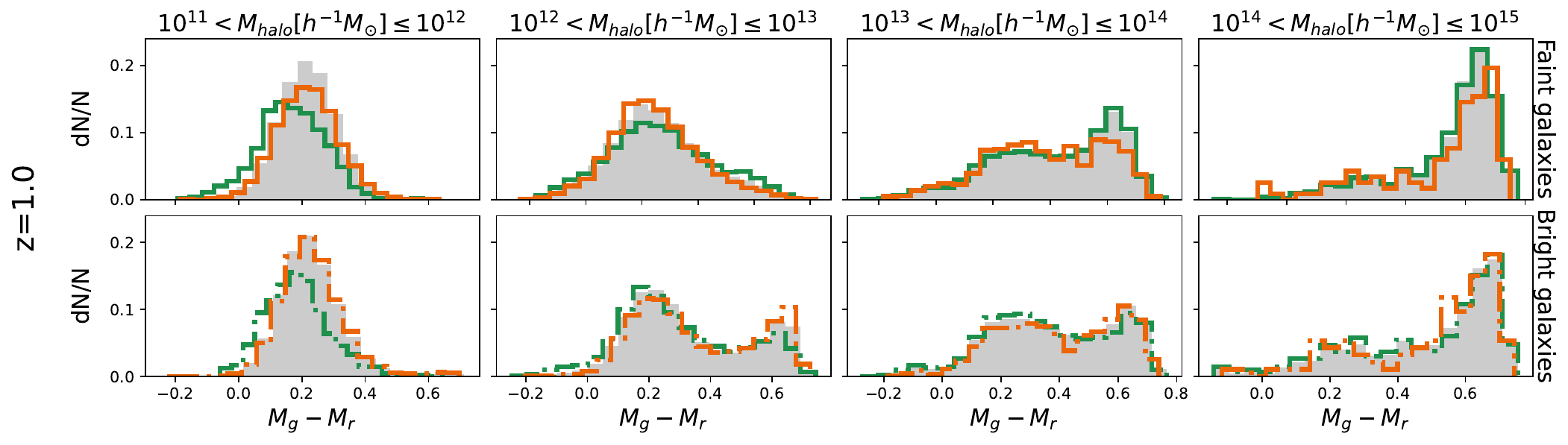}
    }
    {
        \includegraphics[width=0.9\textwidth]{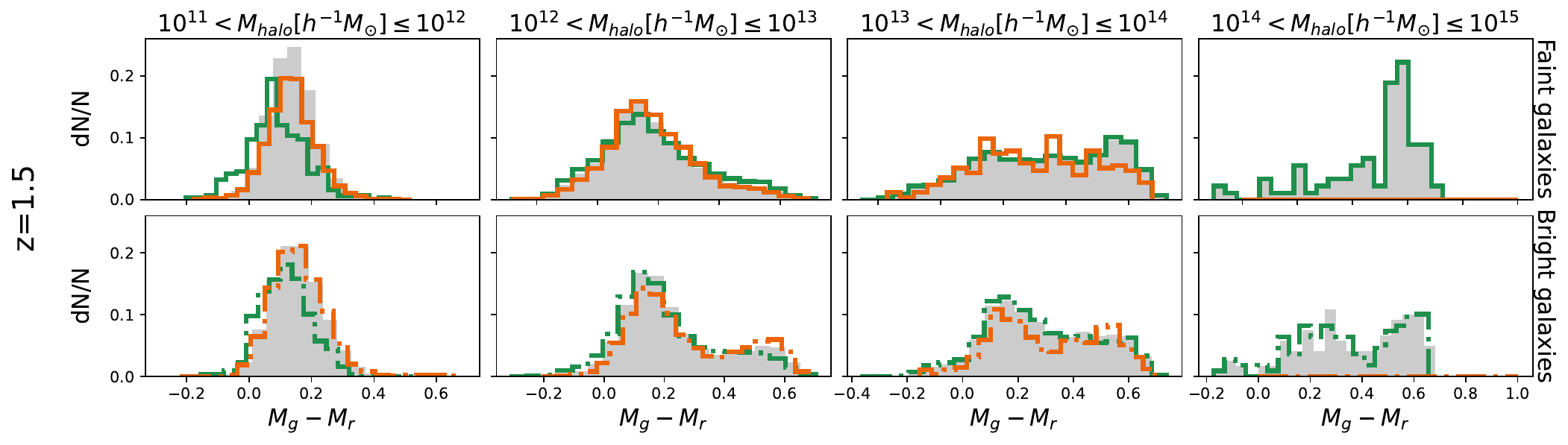}
    }
    {
        \includegraphics[width=0.9\textwidth]{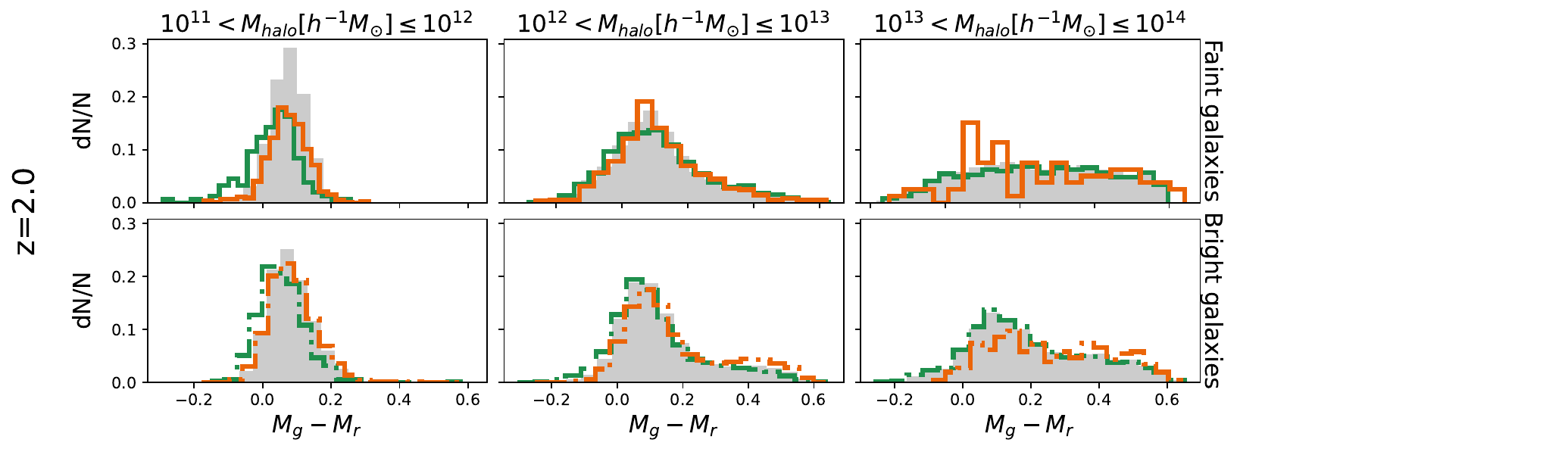}
    }    
    \caption{The colour distribution for faint and bright galaxies belonging to node, filament and total sample. The distributions are shown for each redshift from z=0 (top) to z=2 (bottom) and for mass halo bins (left to right).}
    \label{fig:ev_color}
\end{figure*}

To perform a more detailed analysis, we studied the evolution of the galaxy properties, taking into account colours.
The colour $M_g - M_r$ has been calculated from the SubhaloStellarPhotometrics catalogue \citep{Nelson2018}.
Figure \ref{fig:ev_color} shows the colour distributions for faint and bright galaxies at $z=0.0, 0.5, 1.0, 1.5 \ \text{and} \ 2$ belonging to nodes, filaments and total samples.  Furthermore, the galaxy colour distributions are divided according to the host halo mass.
To ensure that the result is not biased by the stellar mass of the subhalos, a subhalo mass-matching procedure was performed.
This allows us to isolate the impact of the nodes or filaments on the observed differences.

In the lower halo mass bin $\left(10^{11} < M_{\text{halo}}[h^{-1} M_{\odot}] < 10^{12}\right)$, the three samples indicate that the faint and bright galaxies at $z=2$ are predominantly blue, with a slow shift towards the green valley at local z.
The distributions corresponding to nodes for both galaxy luminosities are shifted towards bluer colours, in agreement with the results of \cite{Perez2024}, who found an excess of blue galaxies in low mass halos associated with nodes.
Halos with masses in the range $\left[10^{12} h^{-1} M_{\odot} - 10^{13} h^{-1} M_{\odot}\right]$ at $z=2$ contain predominantly blue (faint and bright) galaxies and a few red galaxies.
With z, the evolution of the two types of galaxies is distinguished.
While at $z=1$ the colour distribution of faint galaxies is slightly shifted towards redder colours, the bright galaxies show a pronounced bimodal distribution.
At local z this is more apparent, as bright galaxies are strongly blue, while faint galaxies show a mixed population.

In halos with masses in the range $\left[10^{13} h^{-1} M_{\odot} - 10^{14} h^{-1} M_{\odot}\right]$ at $z=2$ the faint galaxies exhibit a wide range of colours, while the bright galaxies are comprised of two distinct populations, with a higher blue fraction than the red population.
As redshift decreases, the colour behaviour of both galaxy types becomes similar. At $z=1$, the colour distribution is bimodal, while at $z=0$, the halo population is predominantly red with few blue galaxies. It is noticeable that the nodes have a greater number of faint red galaxies than the rest of the samples in this mass halo bin.
The most massive halos at $z=2$ have not yet formed, which is consistent with the findings of Fig. \ref{fig:hod}.
At $z=1.5$, it is observed that these massive halos are only present in node regions and with a few galaxies whose colours span a wide range. This result is consistent with the HOD mass ranges shown in Fig. \ref{fig:hod}.
At $z = 1$, the massive halos are populated by both structures and have a predominantly red galaxy population with few blue galaxies. The colour evolution is rapid in these halos, which contain only red galaxies at local z.
Our findings indicate that reddening is a consequence of time evolution, with bright galaxies and massive halo environments exhibiting enhanced and accelerated reddening.

Reddening occurs in all galaxies as part of the gradual evolutionary process. In addition, some galaxies cease star formation almost completely through a process called quenching, which is driven by ram pressure stripping, active galactic nuclei (AGN) feedback, supernova feedback and strangulation \citep{Peng2012}.
This is when the galaxy exhausts its fuel and stops forming stars for a period of time long enough for the blue population to join the red population \citep{Peng2015}. However, it is still unclear which is the main mechanism responsible for this, or whether several mechanisms occur simultaneously.
These processes seem to be more efficient in high mass halos with a strong effect on bright galaxies.
Both results could be explained by considering that there are two different effects operating, mass quenching and environmental quenching, which are independent of each other \citep{Peng2010}, both of which are important in determining the fraction of red galaxies in a population \citep{Baldry2006}.

\section{Summary and Discussion}\label{sec:conclusions}
In this paper we have performed an study of the evolution of the Halo Occupation Distribution (HOD) and galaxy properties for the following redshift snapshots:  $z=0.0, 0.5, 1, 1.5 \ \text{and} \ 2$ of \textsc{Illustris TNG300-1} simulations.
We use the Cosmic Web Distances catalogue,
based on the \textsc{DisPerSE} code, in order to obtain the cosmic web distances between each subhalo and the nearest cosmic web structures.
We define nodes as halos with distance to the node $\text{d}_{\text{n}} \leq 1\ \text{R}_{\text{200}}$ and filaments as halos with distance to the node $\text{d}_{\text{n}} > 1\ \text{R}_{\text{200}}$ and 
within an overdensity region $\delta + 1 = 10$ of the radial density profile.

We selected subhalos with masses $\text{M}_* > 10^{8.5}\ h^{-1} \text{M}_{\odot}$ belonging to halos with masses $\text{M}_{200} \geq 10^{11}\ h^{-1} \text{M}_{\odot}$.
We found traces of the flow of galaxies in the cosmic web, obtaining an increase of $15\%$ in the fraction of galaxies in the nodes and a decrease of $3\%$ in the filaments as $z$ decreases.
The fraction of halos in both structures shows slight variations, which can be related to the redshift range considered \citep{Hahn2007}.

We computed the halo occupation distribution (HOD) for node and filament samples at different redshift snapshots, considering galaxies with magnitude thresholds: $\text{M}_\text{r} - 5\text{log}_{10}(h) \leq -17, -18, -19\ \text{and} -20$.
We found that the halo occupancy decreases as the magnitude threshold increases, and the distributions cover an increasing range of halo masses as $z$ decreases.
Also, halos require more initial mass to host bright galaxies than faint ones.

Regarding to the nodes, there are differences only at the low mass extremes for faint galaxies, while for bright galaxies there are differences across the mass range.
This seems to indicate that as redshift approaches zero, the number of faint galaxies increases in the low mass halos, while no significant differences are seen in high mass halos. 
The brighter galaxies, on the other hand, allow us to see that halos increase in mass more than the number of bright galaxies they accrete. 
Therefore, the HOD moves almost in parallel towards higher masses.

For filaments, in contrast to nodes, no large differences in HOD are found for faint galaxies. For brighter galaxies, a similar but noisier behaviour to that of the nodes is observed, with the HOD being shifted towards higher masses. 
The general behaviour found in these structures seems to indicate that the occupation of halos in these environments is influenced both by the accretion of faint galaxies and by processes within the halos, e.g. the merging of galaxies to form more luminous galaxies \citep{Benavides2020}. 

In order to quantify the evolution, we parametrize the HODs by fitting five parameters that characterise the contributions of the central and satellite galaxies:
The $M_{\rm min}$ parameter decreases with time when the central galaxy is faint and increases when the central galaxy is bright, consistent with the variation of the observed HODs at lower halo masses.
This result suggests that halos are populated by a faint central galaxy earlier than at $z=2$, while they need more mass to host a bright central galaxy.
Furthermore, this parameter suggests that the formation of central galaxies (faint or bright) is not affected when the halo belongs to a  filament, but the node regions influence the formation of the central galaxy.
At $z=2$, halos require above-average mass to host a central galaxy, while at local redshift the halo mass required depends on the galaxy magnitude, being above (below) average if the galaxy is bright (faint).

The $\sigma_{\log M}$ parameter also suggests that, for two environments,
the formation of central galaxies is less related to a specific halo mass range than expected, resulting in a larger stellar mass-halo mass scatter.

For faint satellites, $M_1$ does not vary with $z$, while for bright satellites $M_1$ increases from $z=2$ to $z=0$.
This suggests that the mass required for a halo to host a bright satellite galaxy, in whatever environment it is found, is now greater than it was in the past, while there are no variations with z in the mass required to host a faint satellite.
While the filamentary environment does not appear to affect the initial satellite occupation, the nodes require less mass than average to contain a satellite, regardless of the magnitude of the galaxy.

We find that $M_{\rm cut}$ increases slightly with redshift for faint satellites and decreases for bright satellites.
This suggests that at $z=0$ less mass would be required to host a faint satellite than to host a bright satellite, and the opposite at high $z$.
The sample of nodes and filaments takes values lower than the total sample, suggesting that the minimum mass of halos in these regions to host a satellite galaxy is lower than average.

The parameter $\alpha$ varies slightly with $z$ and is close to 1 for the filament sample, while it is around 0.8 for the nodes.
This suggests that the number of satellite galaxies in the halo increases depending on where the halo is located, with a linear increase for filaments and more smoothing for nodes.

We also calculated the $M_{1}/M_{\rm min}$ ratio and found that at local $z$ the central galaxies populate halos over a wider range of halo masses before being occupied by satellite galaxies than at early times. In addition, the nodes have lower values for this ratio, suggesting that these high-density regions are populated by satellite galaxies much earlier than halos in other environments.

These trends align with the findings of \cite{Contreras2023}, who observed that this quotient reaches a plateau at high redshift and then increases towards local z.
It is important to highlight that their samples were selected by stellar masses and fixed number densities, whereas we have split the sample according to stellar magnitude thresholds and different cosmic environments.
Consequently, the evolutionary behaviour of different galaxy populations remains similar, regardless of the specific physics involved in their formation and the cosmological model. 

While the presence of a satellite galaxy, regardless of its magnitude, is not affected if the halo is within a filament, it is affected if the halo is within a high-density region such as nodes.
We observe that the relationship between the number of satellite galaxies and the mass of the halo is different in these regions, meaning that halos in these regions are capable of hosting satellite galaxies with masses lower than those required in other environments.
Regarding the presence of the central galaxy, the mass required varies depending on the magnitude of the galaxy.
In general, at $z=0$ less mass is required to host a faint central galaxy than at early times, while the opposite is true for a bright central galaxy, which requires more mass than at high $z$.
In addition, we find differences at high $z$ for hosting central galaxies when the halo is located at the nodes, suggesting that the distribution and properties of halos in different environments differ significantly at early times.

We estimate the SHMR by separating galaxies into faint and bright ones, and follow their evolution to detect possible variations in both environments.
Filament environments do not appear to influence the stellar mass content of galaxies-whether faint or bright-within the observed scatter. 
However, the influence of nodes on galaxies belonging to halos with masses lower than $10^{12.5} h^{-1} M_{\odot}$ is observable at the local redshifts for faint galaxies and at high redshifts for bright galaxies.
Furthermore, the relation between stellar mass and halo mass depends on the galaxy magnitude.
For faint galaxies, stellar mass increases monotonically with halo mass, whereas for bright galaxies it depends on halo mass.
Below $10^{12.5} h^{-1} M_{\odot}$, the increase in stellar mass is pronounced, and for massive halos the relationship is modest.

Additionally, we performed a comparative analysis between the SHMR and observational data points extracted from \cite{Shuntov2022}.
The findings indicate that this relation exhibits significantly higher values in the observational data set.
The discrepancy may be attributed to variations in the methodologies employed to estimate galaxy stellar masses and magnitudes, as well as the star formation models utilised. 
Moreover, comparing simulations with observational surveys that cover relatively small volumes can introduce biases. 
Furthermore, uncertainties may arise from several factors, including the estimation of redshift and stellar masses, clustering properties, the radial distribution of dark matter and galaxies within halos, and halo mass definitions.
Our findings align with those of \cite{Moster2010} only at the low mass halo extreme, suggesting that the feedback mechanisms controlling star formation in low mass halos in the simulations are consistent with observations. 
In contrast, at the high mass halo extreme, discrepancies might be attributed to the less efficient environmental quenching of satellites in the simulations. 
Despite the potential discrepancies in structural and property characteristics resulting from the classification techniques employed to ascertain cosmic web structures \citep{Zhu2024}, our findings are consistent with \cite{Darvish2017}. Both structures exert a comparable influence on the SHMR, with only minor discrepancies observed at local redshift concerning faint galaxies.

Finally, we observe the evolution of the galaxy colours in halo mass bins.
Reddening occurs in the evolutionary process of all galaxies, where some galaxies cease star formation almost completely through a process called quenching, which is driven by ram pressure stripping, active galactic nuclei (AGN) feedback, supernova feedback and strangulation \citep{Peng2012}.
We observe that the reddening is faster in bright galaxies of both structures and seems more efficient in high mass halos.
Both results could be explained by considering that there are two different effects operating, mass quenching and environmental quenching.
Furthermore, the reddening affects galaxies in both nodes and filaments in a similar way, although halos in nodes with masses below $10^{12} h^{-1} M_{\odot}$ have an excess of blue galaxies, in accordance with \cite{Perez2024}.

Mass quenching is an environmentally independent process that mainly affects massive galaxies, which we expect to be the brightest \citep{Baldry2006}.
In addition, with increasing stellar mass, the star formation rate decreases \citep{Kauffmann2003,Wetzel2012}.
Massive galaxies are more likely to redden due to different processes that occur during their formation and evolution.
Initially, the galaxy experiences intense star formation in its early stages, so that later its star formation rate declines due to fuel depletion.
It also forms massive stars that quickly exhaust their fuel and undergo supernova explosions, releasing energy and gas into their surroundings.
By compressing nearby gas clouds, the positive feedback can lead to bursts of star formation. However, the energy and momentum released can also heat and disperse the surrounding gas, making star formation difficult as the gas must cool and condense before it can collapse to form stars.
Moreover, some galaxies with supermassive black holes can have activity that produces active galactic nuclei, whose energy can heat the surrounding gas, preventing it from cooling and forming stars. 
AGN can also reduce the availability of gas for star formation by driving outflows of gas and particles from the centre of the galaxy.

On the other hand, high density within nodes favours environmental quenching, since star formation history is the most environmentally sensitive property of galaxies \citep{Kauffmann2004,Baldry2006}.
The transition of galaxies from the blue to the red sequence is accelerated by the combination of different processes in denser environments.
In addition, galaxies move through the intracluster medium, which is affected by the ram pressure, where gas is gradually removed by the pressure exerted by the environment. 
Moreover, AGNs, which also favour the cessation of star formation, are more likely to be found in denser environments. 
This variation in colour is accompanied by morphological changes that occur more slowly because they are driven by different mechanisms.
Morphological transitions require complex dynamical processes, including mergers, interactions and gravitational instabilities, which operate on long timescales, while bursts of star formation or quenching can rapidly affect the colour \citep{roger3,Ruiz2023}.

The analysis in this paper shows that at local redshift, nodes have an important effect on the halo-galaxy relationship.
The halos associated with nodes acquire their first faint central galaxy earlier, i.e. at a lower mass than the filaments, so the number of galaxies in the low mass halos is higher.
Once the central galaxy has formed, the host of faint satellite galaxies statistically appears much earlier than in halos in filaments.
In addition, nodes have an effect on halos with masses less than $10^{12} h^{-1} M_{\odot}$, which have an excess of blue galaxies with lower-than-average stellar masses.
In these halos, stellar mass is mainly due to star formation, which may be inhibited by ongoing interactions within these still-forming regions.
A comprehensive analysis of the trajectories and evolution of both central and satellite galaxies will provide insights into the events occurring within the nodes. This is a crucial step in our upcoming research.

\section*{Acknowledgements}
We thank the referee, for providing us with helpful comments and suggestions that improved this paper.
This work was supported in part by the Consejo Nacional de Investigaciones Cient\'ificas y T\'ecnicas de la Rep\'ublica Argentina (CONICET) and the Consejo Nacional de Investigaciones Cient\'ificas, T\'ecnicas y de Creaci\'on Art\'istica de la Universidad Nacional de San Juan (CICITCA).
The authors would like to thank the \textsc{IllustrisTNG} team for making their data available to the public.
FR would like to acknowledge support from the ICTP through the Junior Associates Programme 2023-2028.

\section*{Data Availability}
The data underlying this article will be shared on reasonable
request to the corresponding author.
 



\bibliographystyle{mnras}
\bibliography{example} 



\bsp	
\label{lastpage}
\end{document}